\documentclass[aps,prb,reprint,superscriptaddress,nofootinbib]{revtex4-2}
\usepackage{graphicx,epsfig}
\usepackage{times}
\usepackage{graphics,dcolumn,bm,float}
\usepackage{amssymb,amsmath,rotate,color,dsfont}
\usepackage[breaklinks,colorlinks=true,urlcolor=blue,citecolor=blue,linkcolor=blue]{hyperref}
\usepackage[T1]{fontenc}

\begin{document}
\unitlength 1 cm
\newcommand{\be}{\begin{equation}}
\newcommand{\ee}{\end{equation}}
\newcommand{\bearr}{\begin{eqnarray}}
\newcommand{\eearr}{\end{eqnarray}}
\newcommand{\nn}{\nonumber}
\newcommand{\vpdag}{{\vphantom{\dagger}}}
\newcommand{\vecr}{\vec{r}}
\newcommand{\bs}{\boldsymbol}
\newcommand{\up}{\uparrow}
\newcommand{\down}{\downarrow}
\newcommand{\fns}{\footnotesize}
\newcommand{\ns}{\normalsize}
\newcommand{\cdag}{c^{\dagger}}
\newcommand{\so}{\lambda_{\rm SO}}
\newcommand{\jh}{J_{\rm H}}
\newcommand{\tn}{T_{\rm N}}
\newcommand{\tq}{T_{q}}
\newcommand{\la}{\langle}
\newcommand{\ra}{\rangle}
\newcommand{\sgn}{\text{sgn}}

\definecolor{red}{rgb}{1.0,0.0,0.0}
\definecolor{green}{rgb}{0.0,1.0,0.0}
\definecolor{blue}{rgb}{0.0,0.0,1.0}

\newcommand{\red}[1]{\textcolor{red}{#1}}
\newcommand{\violet}[1]{\textcolor{violet}{#1}}
\newcommand{\blue}[1]{\textcolor{blue}{#1}}

\title{High-Temperature Quantum Anomalous Hall Effect in Buckled Honeycomb Antiferromagnets}

\author{Mohsen Hafez-Torbati}
\email{m.hafeztorbati@gmail.com}
\affiliation{Department of Physics, Shahid Beheshti University, 1983969411 Tehran, Iran}

\author{G\"otz S. Uhrig}
\email{goetz.uhrig@tu-dortmund.de}
\affiliation{Condensed Matter Theory, Department of Physics, TU Dortmund University, 44221 Dortmund, Germany}

\begin{abstract}
We propose N\'eel antiferromagnetic (AF) Mott insulators with a buckled honeycomb structure
as potential candidates to host a high-temperature AF Chern insulator (AFCI). Using a generalized Kondo lattice model
we show that the staggered potential
 induced by a perpendicular electric field due to the buckling can drive the AF Mott insulator to an AFCI phase.
We address the temperature evolution of the
Hall conductance and the chiral edge states.
The quantization temperature $T_q$, below which the Hall conductance is quantized, depends essentially on the
strength of the spin-orbit coupling and the hopping parameter, independent of the specific details of the model.
The deviation of the Hall conductance from the quantized value $e^2/h$
above $T_q$ is found to be accompanied by a spectral broadening of the chiral edge states,
reflecting a finite life-time, i.e., a decay.
Using parameters typical for heavy transition-metal elements we predict that the AFCI can survive up to room temperature.
We suggest Sr$_3$CaOs$_2$O$_9$ as a potential compound to realize a high-$T$ AFCI phase.

\end{abstract}

\maketitle

{\it Introduction.} Due to its far-reaching potential applications in topological quantum computation and
low-energy-consumption spintronic devices the Chern insulator (CI) state
has attracted considerable attention in the past decade \cite{Chang2023,Tokura2019}.
This has led to the discovery of the CI in different classes of systems including
thin films of magnetically doped topological insulators \cite{Chang2013,Chang2015,Mogi2015},
thin films of the intrinsic magnetic topological insulator MnBi$_2$Te$_4$ \cite{Zhu2025,Deng2020,Liu2020},
and moir\'{e} materials \cite{Li2021,Serlin2020}.
Despite this remarkable progress, the observation of the CI is limited to temperatures of only a
few Kelvins, arising from the material's negligible charge gap or low magnetic transition temperature.

While the current realization of the CI is limited to ferromagnets, antiferromagnets
are far more common and exhibit generally higher transition temperatures, reaching hundreds of Kelvins \cite{Baltz2018}.
In addition, the AF ordering of strongly correlated electrons
is known
to be accompanied with a noticeable blue shift of the charge gap
\cite{Hafez-Torbati2021,Hafez-Torbati2022,Bossini2020,Sangiovanni2006}.
This is to be compared with the ferromagnetic ordering on the top and the bottom surfaces of MnBi$_2$Te$_4$
inducing almost no gap in the Dirac states \cite{Hao2019,Chen2019,Li2019b,Swatek2020}.
Thus, there is compelling reason to search
for a high-temperature CI in AF Mott materials.

A non-zero Chern number necessitates the time-reversal symmetry to be truly broken \cite{Jiang2018,Ebrahimkhas2022}.
This is inherent in ferromagnets but not guaranteed in antiferromagnets.
The effect of the time-reversal transformation
on an AF state can be compensated by a lattice group operation. This composite anti-unitary symmetry needs to be broken
for the emergence of an AFCI, which is usually achieved by inducing a staggered potential between the spin-up and the spin-down
sublattices \cite{Jiang2018,Ebrahimkhas2022,Guo2023}.
The direction of the magnetization on the higher-energy (or the lower-energy) sublattice determines the clockwise or
the counter-clockwise propagation direction of the chiral edge states.
The AFCI phase is already predicted in various systems
\cite{Jiang2018,Ebrahimkhas2022,Guo2023,Hafez-Torbati2024a,Hafez-Torbati2024,Wu2023}.
For those involving heavy transition-metal elements a large charge gap is also reported \cite{Hafez-Torbati2024,Wu2023}.
However, an explicit study of whether the AFCI can persist up to high temperatures
remains a crucial open question.

Here, we investigate the N\'eel AF transition-metal compounds possessing a buckled honeycomb structure,
see Fig. \ref{eq:heis}(a).
Using a generalized Kondo lattice model we confirm that a perpendicular electric field can drive the AF Mott insulating
state to an AFCI phase. We address
the temperature evolution of the Hall conductance and the chiral edge states.
The quantization temperature $T_q$, below which the Hall conductance is quantized,
shows generic behavior depending only on the spin-orbit coupling and the hopping parameter.
The deviation of the Hall conductance from the quantized value $e^2/h$ above $T_q$
is accompanied by a spectral broadening of the chiral edge modes.
For heavy transition-metal elements we estimate that $T_q$ can reach room temperature.
We propose Sr$_3$CaOs$_2$O$_9$ as a potential candidate to realize a high-$T$ AFCI state.

{\it Model Hamiltonian.} The low-energy properties of transition-metal compounds are commonly described by
the spin-$\mathcal{S}_{\rm tot}$ Heisenberg model \cite{Fazekas1999}
\be
H=J\sum_{\la i,j \ra } \vec{\mathcal{S}}_i \cdot \vec{\mathcal{S}}_j +\cdots \ ,
\label{eq:heis}
\ee
where $\la i,j \ra$ limits $i$ and $j$ to be nearest-neighbor (NN) and the dots stand for terms beyond the
isotropic NN interaction. For the N\'eel AF order,
the NN AF interaction  ($J>0$)
is usually the dominant term.

\begin{figure}[t]
   \begin{center}
   \includegraphics[width=0.42\textwidth,angle=0]{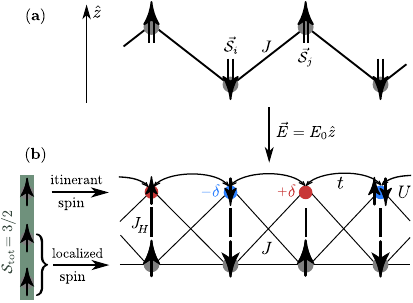}
   \caption{(a) Side view of the spin-$\mathcal{S}_{\rm tot}$ Heisenberg model with the nearest-neighbor AF
   interaction $J$ on the buckled honeycomb structure.
  In the presence of a perpendicular electric field $\vec{E}=E_0\hat{z}$ the system can effectively be described
  by the generalized Kondo lattice model \eqref{eq:model} illustrated in (b) for the special case of
  $\mathcal{S}_{\rm tot}=3/2$.
   }
   \label{fig:model}
   \end{center}
\end{figure}

For the buckled honeycomb structure, see Fig. \ref{fig:model}(a) for a side view, applying a perpendicular electric field
induces a potential difference between the spin-up and the spin-down sublattices.
This induces a shift of charges between the two kinds of sites. The description of
the system requires the Heisenberg model \eqref{eq:heis} to be extended by the charge degrees
of freedom, and the multi-orbital Hubbard model would be the natural choice. However, a reliable analysis
of the topological properties of a strongly correlated multi-orbital model at finite temperature is
a too ambitious numerical challenge. For this reason, we follow the idea proposed in Ref. \cite{Bossini2020}
which was successful in the description of different compounds \cite{Hafez-Torbati2021,Hafez-Torbati2022}.
We approximate the multi-orbital Hubbard model by a generalized Kondo lattice model.
This is well-justified if the charge fluctuations, i.e., the number of holes or double occupancies,
do not exceed one per site.
The generalized Kondo lattice model is given by
(cf. Fig. \ref{fig:model})
\begin{align}
H=&+t\sum_{\langle i,j\rangle} \sum_{\alpha} c^{\dag}_{i\alpha} c^{\vpdag}_{j\alpha}
-2J_{\rm H} \sum_{i} \vec{s}_i \cdot \vec{S}_i
+U \sum_{i} n_{i,\up} n_{i,\down} \nn \\
&+J\sum_{\langle i,j\rangle} (\vec{{S}}_i \cdot \vec{{S}}_j
+\vec{{S}}_i \cdot \vec{{s}}_j
+\vec{{s}}_i \cdot \vec{{S}}_j)  \nn \\
&+\sum_{i} \sum_{\alpha} \delta_i^{\vpdag}  n^{\vpdag}_{i\alpha}
+{\rm i}\lambda_{\rm SO} \sum_{[ i,j ]}
\sum_{\alpha \beta}
\nu_{ij}^{\vpdag}
c^{\dag}_{i\alpha} \sigma^{z}_{\alpha\beta} c^{\vpdag}_{j\beta} \ ,
\label{eq:model}
\end{align}
where $c^{\dag}_{i\alpha}$ is the fermion creation operator at the site $i$ with the $z$-component of the spin
$\alpha=\uparrow$ or $\downarrow$.
The first term is the NN hopping and the second term is the Hund coupling between the electron spin $\vec{s}_i$
and the localized spin $\vec{S}_i$ with the spin quantum number $S=\mathcal{S}_{\rm tot}-1/2$.
We treat one orbital as representative of the charge motion, while the others represent the localized spin.
We emphasize that the Kondo lattice approximation in Eq. \eqref{eq:model} does not mean that we
distinguish between the different orbitals of the same shape. The itinerant orbital stands as a
{\it representative} for all the orbitals \cite{Bossini2020,Hafez-Torbati2021,Hafez-Torbati2022}.
This restricts the possible amount of charge transfer between the higher and the lower energy sublattices.
But, it is very well justified for the values of $\delta$ accessible via realistic strengths of electric fields,
see below and also the ionicity discussed in Supplemental Materials \cite{sm}.

The third term in Eq. \eqref{eq:model} is the Hubbard interaction with $n^{\vpdag}_{i\alpha}:=c^{\dag}_{i\alpha}c^{\vpdag}_{i\alpha}$,
and the fourth term is the Heisenberg interaction between the NN spins.
We always count the Heisenberg interaction on each lattice bond only once.
The fifth term is the staggered sublattice potential giving the onsite energies $+\delta$ and $-\delta$ to the
two sublattices of the honeycomb structure. Figure \ref{fig:model}(b) illustrates these different terms for
the special case of $\mathcal{S}_{\rm tot}=3/2$.

The last term in Eq. \eqref{eq:model} is included to account for the effect of the spin-orbit coupling.
It is exactly the Kane-Mele term \cite{Kane2005b}.
The notation $\left[ i,j\right]$ restricts $i$ and $j$ to be next-nearest-neighbor,
$\sigma^{z}$ stands for the Pauli matrix, and $\nu_{ij}=2/\sqrt{3}(\hat{d}_1 \times \hat{d}_2)_{z}=\pm 1$ where
$\hat{d}_1$ and $\hat{d}_2$ are the unit vectors along the two bonds the electron traverses from site $j$
to site $i$. We add a chemical potential $\mu=U/2$ to the Hamiltonian \eqref{eq:model} to satisfy the half-filling
condition.

We always focus on large values of $U$ and $\jh$ corresponding to the Mott regime and fix the Heisenberg interaction
in Eq. \eqref{eq:model} to $J=4t^2/\Delta_0$ with $\Delta_0:=U+2S\jh$, called the bare Mott gap. This guarantees the Heisenberg model
\eqref{eq:heis} as the low-energy effective model of the Hamiltonian \eqref{eq:model} for $\delta=0$,
apart from some weak anisotropic interactions originating from the spin-orbit coupling $\so$. The Hamiltonian
\eqref{eq:model} generalizes the purely spin model \eqref{eq:heis} allowing for the study of the charge fluctuations induced
by a perpendicular electric field.

The number of methods which allow for a reliable investigation of the topological properties of the strongly correlated
Hamiltonian \eqref{eq:model} at finite temperature are rare.
We employ the dynamical mean-field theory (DMFT) \cite{Georges1996} as an established method for strongly
correlated systems and use the exact diagonalization (ED) as the impurity solver \cite{Georges1996,Caffarel1994}.
We specifically use the real-space realization of the DMFT \cite{Potthoff1999,Song2008,Snoek2008} providing access to the bulk and the edge
properties on equal footing \cite{Hafez-Torbati2018}.
The lattice model is mapped to a set of effective Anderson-Kondo impurity models, which treat the localized spin
quantum mechanically \cite{Hafez-Torbati2021,Hafez-Torbati2022}.
The method takes the local charge and spin quantum fluctuations fully into account.
The local quantum fluctuations leading to a frequency-dependent self-energy are essential for the description of
the paramagnetic Mott insulator at finite temperature. The non-local Heisenberg interactions are replaced by
an effective AF field via a mean-field decoupling \cite{Mueller-Hartmann1989}. The strength of the effective AF field is determined self-consistently
during the DMFT loop \cite{Hafez-Torbati2021,Hafez-Torbati2022}. Further technical details can be found in
Supplemental Materials \cite{sm}. Previous studies of similar interacting topological models by different methods indicate only
minor effects to the phase boundaries due to the non-local quantum fluctuations, neglected in our investigation
\cite{Vanhala2016,Mertz2019,Tupitsyn2019,He2024,Hafez-Torbati2025}.
We will compare the results for the different number of bath sites in the ED calculations.

The Hall conductance for an interacting model at finite temperature can be computed \cite{Ishikawa1987,Yoshida2012,Irsigler2021}
from the
Matsubara Green's function
$\boldsymbol{G}^{\vpdag}_\alpha({\rm i}\omega_n,\vec{k})=
\left[ {\rm i}\omega_n \boldsymbol{\mathds{1}}-\boldsymbol{\mathcal{H}}^{(0)}_\alpha(\vec{k})-\boldsymbol{\Sigma}^{\vpdag}_\alpha({\rm i}\omega_n) \right]^{-1}$ as
\begin{align}
\label{eq:hall}
\sigma_{xy}=\frac{e^2}{h}\frac{T}{12\pi} \epsilon_{\mu\nu\rho}~
{\rm Im}\!\left[ \sum_{\alpha} \sum_n \int \!{\rm d}\vec{k}
~{\rm Tr}\!\left[
\boldsymbol{G}_\alpha \partial_\mu \boldsymbol{G}_\alpha^{-1}
\right.
\right. \nn \\
\left.
\left.
\times
\boldsymbol{G}_\alpha \partial_\nu \boldsymbol{G}_\alpha^{-1}
\boldsymbol{G}_\alpha \partial_\rho \boldsymbol{G}_\alpha^{-1}
\right]
\vphantom{\sum_\alpha\int}
\right]
\end{align}
where $\boldsymbol{\mathcal{H}}^{(0)}_\alpha(\vec{k})$ denotes the Bloch Hamiltonian matrix, $\boldsymbol{G}^{\vpdag}_\alpha \equiv \boldsymbol{G}^{\vpdag}_\alpha({\rm i}\omega_n,\vec{k})$,
$\epsilon_{\mu\nu\rho}$ is the totally antisymmetric tensor, and summations are implied over the indices
$\mu$, $\nu$, and $\rho$, each of which runs over $\mathrm{i}\omega_n$, $k_x$, and $k_y$ \cite{sm}.

The momentum-resolved spectral function is accessed using the real-frequency Green's function,
\be
A_{\alpha,\vec{d}}~(\omega,\vec{k})=
-{\frac{1}{\pi}\rm Im}\! \left[\boldsymbol{G}^{\vpdag}_\alpha(\omega+{\rm i}\eta,\vec{k})\right]_{\vec{d},\vec{d}}
\label{eq:spectral}
\ee
where $\vec{d}$ specifies a lattice site in the unit cell. The broadening factor $\eta=0.01t$
is used in the calculations. The local spectral function $A_{\alpha,\vec{d}}~(\omega)$
is computed by integrating over the momentum with the appropriate prefactor. The spectral sum rule
is always satisfied with an error less than $10^{-3}$ in all the results that we present.

\begin{figure}[t]
   \begin{center}
   \includegraphics[width=0.33\textwidth,angle=-90]{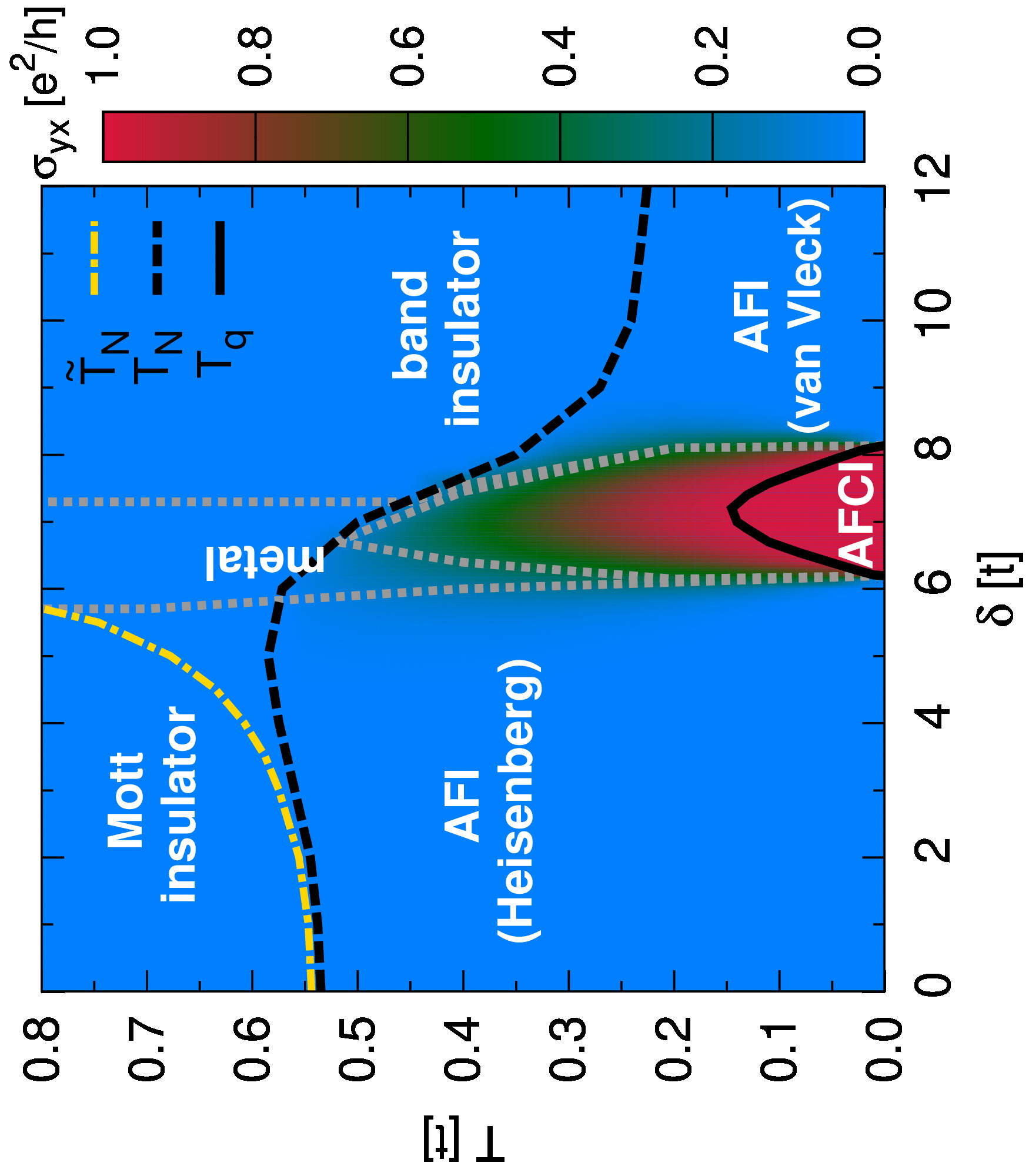}
   \caption{Phase diagram of temperature vs the alternating sublattice potential controlled by a
   perpendicular electric field.
   The colormap represents the value of the Hall conductance $\sigma_{yx}=-\sigma_{xy}$ given
   by Eq. \eqref{eq:hall}. The N\'eel temperature $\tn$ and the crossover quantization temperature $\tq$
   are specified. The Hall conductance takes the quantized value $e^2/h$ with an error less than
   $\%1$ below $\tq$, characterizing the antiferromagnetic Chern insulator (AFCI).
   The gray dotted lines separate the metallic region from the insulating regions.
   The metallic region
   shrinks rapidly to the two quantum critical points as $T\to 0$.
   The mean-field N\'eel temperature $\tilde{T}_{\rm N}$ of the effective low-energy spin model valid for small values
   of $\delta$, see the main text, is also included.
   The results are for the model parameters $S=1/2$, $U=12t$, $\jh=0.2U$,
   $J=4t^2/\Delta_0=0.2\bar{7}t$, and $\so=0.2t$.
   We recall that the localized spin $S=1/2$ in Eq. \eqref{eq:model} corresponds to the total spin
   $\mathcal{S}_{\rm tot}=S+1/2=1$ in Eq. \eqref{eq:heis}.
   The number of bath sites $n_b=5$ is used
   in the ED impurity solver.}
   \label{fig:pd}
   \end{center}
\end{figure}

\begin{figure}[t]
   \begin{center}
   \includegraphics[width=0.3\textwidth,angle=-90]{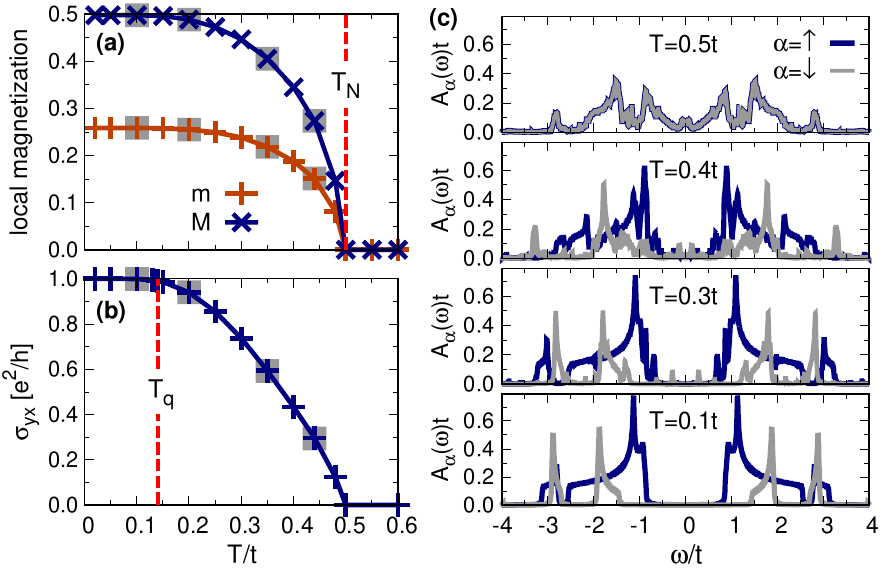}
   \caption{Local magnetizations $m=|\la s_i^z \ra|$ and $M=|\la S_i^z \ra|$ (a) and the
   Hall conductance (b) vs $T$. The N\'eel temperature $\tn$ and the crossover quantization temperature $\tq$ are
   specified. (c) Local spectral function averaged over the two sites in the unit cell for the spin
   component $\alpha$ at different temperatures.
   The results correspond to the solution with the magnetization on the higher-energy sublattice pointing in the
   positive $z$ direction.
   The results are for $S=1/2$, $U=12t$, $\jh=0.2U$,
   $J=4t^2/\Delta_0=0.2\bar{7}t$, $\so=0.2t$, and $\delta=7t$. The data are for the number of bath sites $n_b=6$
   except for the filled gray squares at selective temperatures in (a) and (b) which are for
   $n_b=7$.}
   \label{fig:del7}
   \end{center}
\end{figure}

{\it Phase diagram.} Figure \ref{fig:pd} represents the phase diagram of temperature vs the
alternating sublattice potential for the model parameters $S=1/2$, $U=12t$, $\jh=0.2U$, $J=4t^2/\Delta_0=0.2\bar{7}t$,
and $\so=0.2t$.
Note that the localized spin $S=1/2$ in Eq. \eqref{eq:model}
corresponds to the total spin $\mathcal{S}_{\rm tot}=1$ in Eq. \eqref{eq:heis}, see Fig. \ref{fig:model}.
The number of bath sites $n_b=5$ is used in the ED. The colormap
displays the value of the Hall conductance $\sigma_{yx}=-\sigma_{xy}$. The N\'eel temperature $\tn$
and the crossover quantization temperature $T_q$ are specified. The Hall conductance acquires the quantized value
$e^2/h$ with an error less than $\%1$ below $T_q$, characterizing the AFCI. The gray dotted lines
separate the metallic region from the insulating regions. The metallic region shrinks rapidly
to the two quantum critical points as $T\to 0$.

\begin{figure}[b]
   \begin{center}
   \includegraphics[width=0.41\textwidth,angle=0]{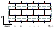}
   \caption{The brick wall representation of the honeycomb structure with the open boundary condition in $x$
   and the periodic boundary condition in $y$ direction. The different sites in the $x$ direction are labeled from
   $0$ to $N_x-1$. The dashed box specifies the unit cell.}
   \label{fig:bw}
   \end{center}
\end{figure}

\begin{figure*}[t]
   \begin{center}
   \includegraphics[width=0.274\textwidth,angle=-90]{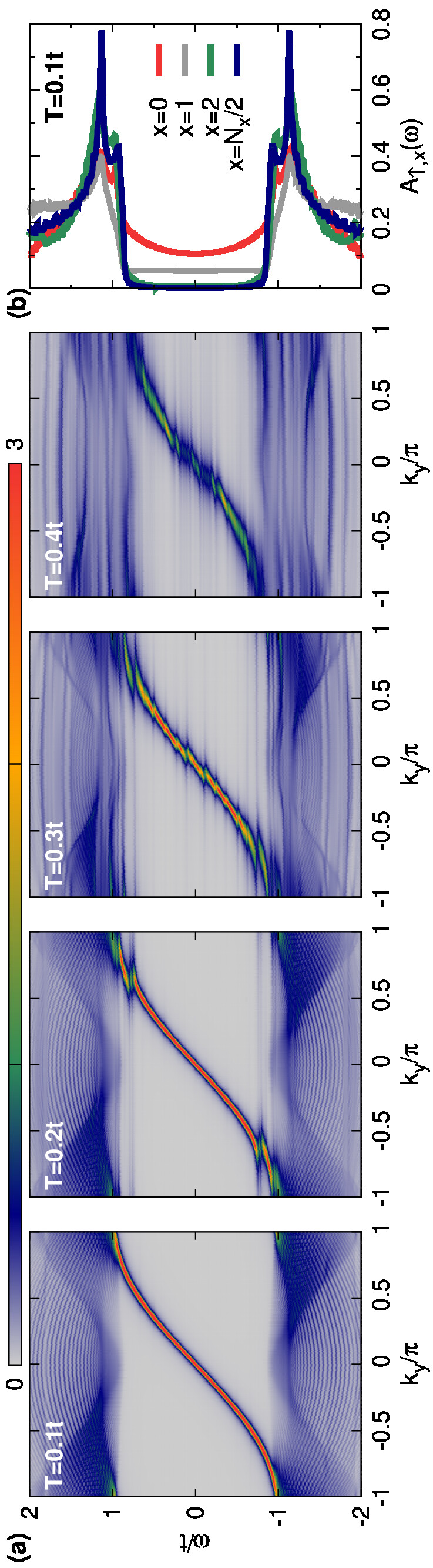}
   \caption{(a) The momentum-resolved spectral function $A_{\alpha,x}(\omega,k_y)$ for the topological spin
   component $\alpha=\up$ at the edge $x=0$ at different temperatures. (b) Local spectral function $A_{\up,x}(\omega)$ for $T=0.1t$
   at different $x$.
   The results are for the same model parameters as in Fig. \ref{fig:del7}.
   A cylindrical geometry as illustrated in Fig. \ref{fig:bw} with $N_x=80$ is used.
   The data are for $n_b=6$ bath sites in the ED impurity solver.}
   \label{fig:spectral}
   \end{center}
\end{figure*}

For small values of $\delta$ the system is a paramagnetic Mott insulator at high and an AFI at low $T$,
which is the well-known physics of the Heisenberg model \cite{Manousakis1991}.
Upon increasing $\delta$, the N\'eel temperature first increases. This is due to the increase in the effective
Heisenberg interaction $J_{\rm eff}=4t^2\Delta_0/(\Delta_0^2-4\delta^2)$ between the {\it itinerant} spins.
The mean-field N\'eel temperature $\tilde{T}_{\rm N}$ of the effective low-energy spin model  \cite{sm} nicely describes
the N\'eel temperature $\tn$ of the generalized Kondo lattice model up to $\delta \simeq 3t$.
We attribute the small deviation to the contributions beyond the second order
perturbation theory used in the derivation of the effective spin model.
For $\delta \gtrsim 5t$ the charge fluctuations start to play a qualitative role in the low-energy physics
causing a drop in $\tn$.
This behavior is reminiscent of the N\'eel temperature of the Hubbard model as the Hubbard interaction
is reduced to lower values than in the Mott regime \cite{Rohringer2018}.
For large values of $\delta$ the lower-energy sublattice is almost doubly occupied and the higher-energy
sublattice is almost empty of electrons. The itinerant electrons show a very weak local magnetization due to the
van Vleck mechanism \cite{Vleck1932,sm}.
Thus, the itinerant electrons are in the band insulator phase above $\tn$.

The intermediate values of $\delta$ are of particular interest because the Hall conductance becomes non-zero and approaches
the quantized value $e^2/h$ as the temperature is lowered.
For $\delta=7t$ we
plot the local magnetizations $m=|\la s_i^z \ra|$ and $M=|\la S_i^z \ra|$ in Fig. \ref{fig:del7}(a)
and the Hall conductance vs $T$ in Fig. \ref{fig:del7}(b). Figure \ref{fig:del7}(c) represents the
local spectral function averaged over the two sites in the unit cell for the spin component $\alpha$
near the Fermi energy $\omega=0$ at different temperatures.
The finite spectral weight developing at the Fermi energy at the paramagnetic high temperatures
signals a metallic state. An intermediate metallic phase separating the band and the Mott insulators
is already reported for the ionic Hubbard model \cite{Garg2006,Paris2007}.
Our results suggest that such a metallic phase appears also in multi-orbital systems at paramagnetic high temperatures.
At low temperatures, the system is expected to be an insulator \cite{Byczuk2009,Kancharla2007,Hafez-Torbati2016}.

The data in Fig. \ref{fig:del7} are for $n_b=6$ bath sites except for the filled gray squares
at selective temperatures in panels (a) and (b) which are for $n_b=7$. The results for $n_b=5$, not included in the figure,
also nicely match the results for $n_b=6$ and 7. This corroborates the accuracy of the results.

{\it Temperature evolution of chiral edge states.}
The quantized Hall conductance at intermediate values of $\delta$ in Fig. \ref{fig:pd} is expected
to support chiral edge states. The direction of magnetization on the higher-energy (or the lower-energy)
sublattice determines the clockwise or counter-clockwise propagation direction of chiral edge states.
We focus on the solution with the magnetization on the higher-energy sublattice pointing in the positive
$z$ direction. The other solution can be accessed using the time-reversal transformation.

To investigate the presence and the temperature evolution of
the chiral edge states we consider a cylindrical geometry with open boundary condition in the $x$ direction
and edges of type armchair. The honeycomb structure is treated as a brick wall labeling the lattice sites
in the $x$ direction from $0$ to $N_x-1$, as illustrated in Fig. \ref{fig:bw}. At each $x$, there are two nonequivalent
lattice sites in the $y$ direction. We consider $N_x=80$.

The momentum-resolved spectral function $A_{\up,x=0}(\omega,k_y)$,
averaged over the two nonequivalent lattice sites in the $y$ direction,
is plotted in Fig. \ref{fig:spectral}(a) for the same model parameters
as in Fig. \ref{fig:del7}.
The results are for the topological spin component $\alpha=\up$.
The spin component $\alpha=\down$ has no contribution to the
quantized Hall conductance at low-$T$ in Fig. \ref{fig:del7}(b) and we
find it to be gapped in the bulk and at the edges.
The chiral edge state in Fig. \ref{fig:spectral}(a) at $T=0.1t$ quickly disappears as the bulk is approached.
This can be seen from the momentum-integrated spectral function $A_{\up,x}(\omega)$
given for different values of $x$ in Fig. \ref{fig:spectral}(b).
The temperature evolution of the momentum-resolved spectral function in Fig. \ref{fig:spectral}(a)
indicates a spectral broadening as the temperature is increased above $T \sim 0.2t$.
This reflects a finite life-time, i.e., a decay, of the chiral edge state and is in accord with the deviation
of the Hall conductance from the quantized value $e^2/h$ above $T_q$
in Fig. \ref{fig:del7}(b).

\begin{figure}[t]
   \begin{center}
   \includegraphics[width=0.33\textwidth,angle=-90]{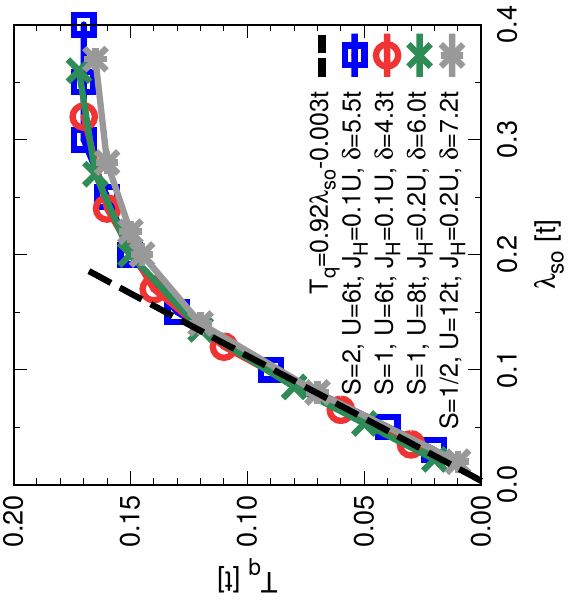}
   \caption{Quantization temperature of the Hall conductance vs the spin-orbit coupling for
   different sets of model parameters. The Heisenberg interaction is always fixed to $J=4t^2/\Delta_0$.
   In each case, the optimal value of $\delta$ is chosen such that the
   system is in the middle of the AFCI phase.
    The results are for the number of bath sites $n_b=5$. The dashed line represents a linear fit to the data for $\so<0.15t$.}
   \label{fig:tq}
   \end{center}
\end{figure}

{\it Quantization temperature.}
Up to now, our study has been limited only to particular model parameters. Nevertheless,
the AFCI always appears when  the sublattice potential difference $2\delta$ is
of the same order as the bare Mott gap $\Delta_0=U+2S\jh$, independent of the details of the model.
The quantization temperature $T_q$ also indicates generic behavior. To demonstrate this,
we have plotted $T_q$ vs the spin-orbit coupling in Fig. \ref{fig:tq} for different sets
of model parameters. In each case, the optimal value of $\delta$ is chosen such that the system is in the middle
of the AFCI phase, see Fig. \ref{fig:pd} for $\delta=7.2t$ used in the last set of parameters.
The Heisenberg interaction is always fixed to $J=4t^2/\Delta_0$.

Interestingly, despite the very different model parameters used in Fig. \ref{fig:tq} one can hardly see any change in $T_q$.
The quantization temperature is solely determined by the spin-orbit coupling and the hopping parameter.
$T_q$ increases linearly with $\so$ and saturates for large values of $\so$.
Although the spin-orbit coupling is highly compound-specific, one usually expect $\so<0.15t$.
In this region, the data can nicely be described
by the linear fit $T_q=0.92\so-0.003t$ shown in Fig. \ref{fig:tq} as a black dashed line.
This relation allows us to estimate the scale of the quantization temperature
based on the knowledge of the spin-orbit coupling and the hopping parameter.

The precise determination of the strength of the spin-orbit coupling $\so$ in a material is
challenging. However, it is usually estimated to be of the order of a few meV for $3d$, around $15$ meV for
$4d$, and around $40$ meV for $5d$ transition-metal elements \cite{Khomskii2021}. Note that the numbers refer
to the Kane-Mele-type of the spin-orbit coupling. For comparison, the precise value of $\so$ for silicene,
germanene, and stanene are computed to be $0.77$, $8.9$, and $12$ meV, respectively \cite{Liu2011}.
Assuming $t \sim 200$ meV for $3d$ transition-metal compounds one finds that it is possible to
realize the AFCI in these materials although with the small quantization temperature $\tq \sim 10$ K.
This estimate nicely agrees with first principles calculations
predicting the AFCI in CrO with a small charge gap of about $1$ meV at zero temperature \cite{Guo2023}.
For the more extended $4d$ orbital with the typical hopping parameter
$t \sim 400$ meV we find $\tq \sim 100$ K. Assuming $t \sim 700$ meV for $5d$ materials,
our results unveil the possible realization of the AFCI at room temperature.

{\it Conclusion.} We propose the strongly correlated AF materials with a buckled
honeycomb structure as potential candidates to realize a hight-$T$ CI phase.
The system can be driven from the AF Mott insulating state to the AFCI phase by
applying a perpendicular electric field, which allows to induce and fine-tune a sublattice potential
difference. The AFCI emerges when the sublattice potential difference reaches the same size as the Mott gap.
The experimentally accessible electric field of $0.5$ V/\r{A} \cite{Ni2012,Zhang2009a,Sakanashi2021}
can induce a sublattice potential
difference of a few eV in a buckling height of a few angstroms. Since the Mott gap and the buckling
height can also be effectively controlled by pressure, we believe the AFCI state is realistically within reach.

Compounds Ba$_2$NiTeO$_6$ and Sr$_3$CaOs$_2$O$_9$ are known buckled honeycomb antiferromagnets with
the buckling heights of $5.3$ \r{A} \cite{Asai2016,Asai2017} and $3.3$ \r{A} \cite{Thakur2022}, respectively.
The latter compound shows a N\'eel AF order with the high N\'eel temperature $\tn \sim 385$ K and is
a promising candidate to host a high-$T$ AFCI.
It should be noted that Sr$_3$CaOs$_2$O$_9$ is not anisotropic enough to be considered effectively
two-dimensional and thin-film growth would be needed. It is also conceivable to strain-engineer the
films to enhance the in-plane and further suppress the out-of-plane couplings to put the system closer to
the two-dimensional limit.
Growing $[111]$-oriented bilayers of perovskite heavy transition-metal oxides is another route to
a buckled honeycomb antiferromagnet \cite{Xiao2011} and a high-$T$ AFCI state.
The possible realization of an AFCI in $[111]$-oriented perovskite bilayers has been previously
predicted through studies of static exchange AF fields and first-principles calculations at zero temperature \cite{Liang2013}.
Our analysis going beyond static mean-field approaches and addressing the topological properties at
finite temperatures explicitly unveils that AF Mott insulators can host a quantum anomalous Hall effect that can
persist up to high temperatures.



\begin{thebibliography}{63}%
\makeatletter
\providecommand \@ifxundefined [1]{%
 \@ifx{#1\undefined}
}%
\providecommand \@ifnum [1]{%
 \ifnum #1\expandafter \@firstoftwo
 \else \expandafter \@secondoftwo
 \fi
}%
\providecommand \@ifx [1]{%
 \ifx #1\expandafter \@firstoftwo
 \else \expandafter \@secondoftwo
 \fi
}%
\providecommand \natexlab [1]{#1}%
\providecommand \enquote  [1]{``#1''}%
\providecommand \bibnamefont  [1]{#1}%
\providecommand \bibfnamefont [1]{#1}%
\providecommand \citenamefont [1]{#1}%
\providecommand \href@noop [0]{\@secondoftwo}%
\providecommand \href [0]{\begingroup \@sanitize@url \@href}%
\providecommand \@href[1]{\@@startlink{#1}\@@href}%
\providecommand \@@href[1]{\endgroup#1\@@endlink}%
\providecommand \@sanitize@url [0]{\catcode `\\12\catcode `\$12\catcode
  `\&12\catcode `\#12\catcode `\^12\catcode `\_12\catcode `\%12\relax}%
\providecommand \@@startlink[1]{}%
\providecommand \@@endlink[0]{}%
\providecommand \url  [0]{\begingroup\@sanitize@url \@url }%
\providecommand \@url [1]{\endgroup\@href {#1}{\urlprefix }}%
\providecommand \urlprefix  [0]{URL }%
\providecommand \Eprint [0]{\href }%
\providecommand \doibase [0]{https://doi.org/}%
\providecommand \selectlanguage [0]{\@gobble}%
\providecommand \bibinfo  [0]{\@secondoftwo}%
\providecommand \bibfield  [0]{\@secondoftwo}%
\providecommand \translation [1]{[#1]}%
\providecommand \BibitemOpen [0]{}%
\providecommand \bibitemStop [0]{}%
\providecommand \bibitemNoStop [0]{.\EOS\space}%
\providecommand \EOS [0]{\spacefactor3000\relax}%
\providecommand \BibitemShut  [1]{\csname bibitem#1\endcsname}%
\let\auto@bib@innerbib\@empty
\bibitem [{\citenamefont {Chang}\ \emph {et~al.}(2023)\citenamefont {Chang},
  \citenamefont {Liu},\ and\ \citenamefont {MacDonald}}]{Chang2023}%
  \BibitemOpen
  \bibfield  {author} {\bibinfo {author} {\bibfnamefont {C.-Z.}\ \bibnamefont
  {Chang}}, \bibinfo {author} {\bibfnamefont {C.-X.}\ \bibnamefont {Liu}},\
  and\ \bibinfo {author} {\bibfnamefont {A.~H.}\ \bibnamefont {MacDonald}},\
  }\bibfield  {title} {\bibinfo {title} {Colloquium: Quantum anomalous {H}all
  effect},\ }\href {https://doi.org/10.1103/RevModPhys.95.011002} {\bibfield
  {journal} {\bibinfo  {journal} {Rev. Mod. Phys.}\ }\textbf {\bibinfo {volume}
  {95}},\ \bibinfo {pages} {011002} (\bibinfo {year} {2023})}\BibitemShut
  {NoStop}%
\bibitem [{\citenamefont {Tokura}\ \emph {et~al.}(2019)\citenamefont {Tokura},
  \citenamefont {Yasuda},\ and\ \citenamefont {Tsukazaki}}]{Tokura2019}%
  \BibitemOpen
  \bibfield  {author} {\bibinfo {author} {\bibfnamefont {Y.}~\bibnamefont
  {Tokura}}, \bibinfo {author} {\bibfnamefont {K.}~\bibnamefont {Yasuda}},\
  and\ \bibinfo {author} {\bibfnamefont {A.}~\bibnamefont {Tsukazaki}},\
  }\bibfield  {title} {\bibinfo {title} {Magnetic topological insulators},\
  }\href {https://doi.org/10.1038/s42254-018-0011-5} {\bibfield  {journal}
  {\bibinfo  {journal} {Nature Reviews Physics}\ }\textbf {\bibinfo {volume}
  {1}},\ \bibinfo {pages} {126} (\bibinfo {year} {2019})}\BibitemShut {NoStop}%
\bibitem [{\citenamefont {Chang}\ \emph {et~al.}(2013)\citenamefont {Chang},
  \citenamefont {Zhang}, \citenamefont {Feng}, \citenamefont {Shen},
  \citenamefont {Zhang}, \citenamefont {Guo}, \citenamefont {Li}, \citenamefont
  {Ou}, \citenamefont {Wei}, \citenamefont {Wang}, \citenamefont {Ji},
  \citenamefont {Feng}, \citenamefont {Ji}, \citenamefont {Chen}, \citenamefont
  {Jia}, \citenamefont {Dai}, \citenamefont {Fang}, \citenamefont {Zhang},
  \citenamefont {He}, \citenamefont {Wang}, \citenamefont {Lu}, \citenamefont
  {Ma},\ and\ \citenamefont {Xue}}]{Chang2013}%
  \BibitemOpen
  \bibfield  {author} {\bibinfo {author} {\bibfnamefont {C.-Z.}\ \bibnamefont
  {Chang}}, \bibinfo {author} {\bibfnamefont {J.}~\bibnamefont {Zhang}},
  \bibinfo {author} {\bibfnamefont {X.}~\bibnamefont {Feng}}, \bibinfo {author}
  {\bibfnamefont {J.}~\bibnamefont {Shen}}, \bibinfo {author} {\bibfnamefont
  {Z.}~\bibnamefont {Zhang}}, \bibinfo {author} {\bibfnamefont
  {M.}~\bibnamefont {Guo}}, \bibinfo {author} {\bibfnamefont {K.}~\bibnamefont
  {Li}}, \bibinfo {author} {\bibfnamefont {Y.}~\bibnamefont {Ou}}, \bibinfo
  {author} {\bibfnamefont {P.}~\bibnamefont {Wei}}, \bibinfo {author}
  {\bibfnamefont {L.-L.}\ \bibnamefont {Wang}}, \bibinfo {author}
  {\bibfnamefont {Z.-Q.}\ \bibnamefont {Ji}}, \bibinfo {author} {\bibfnamefont
  {Y.}~\bibnamefont {Feng}}, \bibinfo {author} {\bibfnamefont {S.}~\bibnamefont
  {Ji}}, \bibinfo {author} {\bibfnamefont {X.}~\bibnamefont {Chen}}, \bibinfo
  {author} {\bibfnamefont {J.}~\bibnamefont {Jia}}, \bibinfo {author}
  {\bibfnamefont {X.}~\bibnamefont {Dai}}, \bibinfo {author} {\bibfnamefont
  {Z.}~\bibnamefont {Fang}}, \bibinfo {author} {\bibfnamefont {S.-C.}\
  \bibnamefont {Zhang}}, \bibinfo {author} {\bibfnamefont {K.}~\bibnamefont
  {He}}, \bibinfo {author} {\bibfnamefont {Y.}~\bibnamefont {Wang}}, \bibinfo
  {author} {\bibfnamefont {L.}~\bibnamefont {Lu}}, \bibinfo {author}
  {\bibfnamefont {X.-C.}\ \bibnamefont {Ma}},\ and\ \bibinfo {author}
  {\bibfnamefont {Q.-K.}\ \bibnamefont {Xue}},\ }\bibfield  {title} {\bibinfo
  {title} {Experimental {Observation of the Quantum Anomalous {H}all Effect in
  a Magnetic Topological Insulator}},\ }\href
  {https://doi.org/10.1126/science.1234414} {\bibfield  {journal} {\bibinfo
  {journal} {Science}\ }\textbf {\bibinfo {volume} {340}},\ \bibinfo {pages}
  {167} (\bibinfo {year} {2013})}\BibitemShut {NoStop}%
\bibitem [{\citenamefont {Chang}\ \emph {et~al.}(2015)\citenamefont {Chang},
  \citenamefont {Zhao}, \citenamefont {Kim}, \citenamefont {Zhang},
  \citenamefont {Assaf}, \citenamefont {Heiman}, \citenamefont {Zhang},
  \citenamefont {Liu}, \citenamefont {Chan},\ and\ \citenamefont
  {Moodera}}]{Chang2015}%
  \BibitemOpen
  \bibfield  {author} {\bibinfo {author} {\bibfnamefont {C.-Z.}\ \bibnamefont
  {Chang}}, \bibinfo {author} {\bibfnamefont {W.}~\bibnamefont {Zhao}},
  \bibinfo {author} {\bibfnamefont {D.~Y.}\ \bibnamefont {Kim}}, \bibinfo
  {author} {\bibfnamefont {H.}~\bibnamefont {Zhang}}, \bibinfo {author}
  {\bibfnamefont {B.~A.}\ \bibnamefont {Assaf}}, \bibinfo {author}
  {\bibfnamefont {D.}~\bibnamefont {Heiman}}, \bibinfo {author} {\bibfnamefont
  {S.-C.}\ \bibnamefont {Zhang}}, \bibinfo {author} {\bibfnamefont
  {C.}~\bibnamefont {Liu}}, \bibinfo {author} {\bibfnamefont {M.~H.~W.}\
  \bibnamefont {Chan}},\ and\ \bibinfo {author} {\bibfnamefont {J.~S.}\
  \bibnamefont {Moodera}},\ }\bibfield  {title} {\bibinfo {title}
  {High-precision realization of robust quantum anomalous {Hall} state in a
  hard ferromagnetic topological insulator},\ }\href
  {https://doi.org/10.1038/nmat4204} {\bibfield  {journal} {\bibinfo  {journal}
  {Nature Materials}\ }\textbf {\bibinfo {volume} {14}},\ \bibinfo {pages}
  {473} (\bibinfo {year} {2015})}\BibitemShut {NoStop}%
\bibitem [{\citenamefont {Mogi}\ \emph {et~al.}(2015)\citenamefont {Mogi},
  \citenamefont {Yoshimi}, \citenamefont {Tsukazaki}, \citenamefont {Yasuda},
  \citenamefont {Kozuka}, \citenamefont {Takahashi}, \citenamefont {Kawasaki},\
  and\ \citenamefont {Tokura}}]{Mogi2015}%
  \BibitemOpen
  \bibfield  {author} {\bibinfo {author} {\bibfnamefont {M.}~\bibnamefont
  {Mogi}}, \bibinfo {author} {\bibfnamefont {R.}~\bibnamefont {Yoshimi}},
  \bibinfo {author} {\bibfnamefont {A.}~\bibnamefont {Tsukazaki}}, \bibinfo
  {author} {\bibfnamefont {K.}~\bibnamefont {Yasuda}}, \bibinfo {author}
  {\bibfnamefont {Y.}~\bibnamefont {Kozuka}}, \bibinfo {author} {\bibfnamefont
  {K.~S.}\ \bibnamefont {Takahashi}}, \bibinfo {author} {\bibfnamefont
  {M.}~\bibnamefont {Kawasaki}},\ and\ \bibinfo {author} {\bibfnamefont
  {Y.}~\bibnamefont {Tokura}},\ }\bibfield  {title} {\bibinfo {title} {Magnetic
  modulation doping in topological insulators toward higher-temperature quantum
  anomalous {H}all effect},\ }\href {https://doi.org/10.1063/1.4935075}
  {\bibfield  {journal} {\bibinfo  {journal} {Applied Physics Letters}\
  }\textbf {\bibinfo {volume} {107}},\ \bibinfo {pages} {182401} (\bibinfo
  {year} {2015})}\BibitemShut {NoStop}%
\bibitem [{\citenamefont {Zhu}\ \emph {et~al.}(2025)\citenamefont {Zhu},
  \citenamefont {Feng}, \citenamefont {Zhou}, \citenamefont {Wang},
  \citenamefont {Yao}, \citenamefont {Lian}, \citenamefont {Lin}, \citenamefont
  {He}, \citenamefont {Lin}, \citenamefont {Wang}, \citenamefont {Wang},
  \citenamefont {Yang}, \citenamefont {Li}, \citenamefont {Wu}, \citenamefont
  {Liu}, \citenamefont {Wang}, \citenamefont {Shen}, \citenamefont {Zhang},
  \citenamefont {Wang},\ and\ \citenamefont {Wang}}]{Zhu2025}%
  \BibitemOpen
  \bibfield  {author} {\bibinfo {author} {\bibfnamefont {J.}~\bibnamefont
  {Zhu}}, \bibinfo {author} {\bibfnamefont {Y.}~\bibnamefont {Feng}}, \bibinfo
  {author} {\bibfnamefont {X.}~\bibnamefont {Zhou}}, \bibinfo {author}
  {\bibfnamefont {Y.}~\bibnamefont {Wang}}, \bibinfo {author} {\bibfnamefont
  {H.}~\bibnamefont {Yao}}, \bibinfo {author} {\bibfnamefont {Z.}~\bibnamefont
  {Lian}}, \bibinfo {author} {\bibfnamefont {W.}~\bibnamefont {Lin}}, \bibinfo
  {author} {\bibfnamefont {Q.}~\bibnamefont {He}}, \bibinfo {author}
  {\bibfnamefont {Y.}~\bibnamefont {Lin}}, \bibinfo {author} {\bibfnamefont
  {Y.}~\bibnamefont {Wang}}, \bibinfo {author} {\bibfnamefont {Y.}~\bibnamefont
  {Wang}}, \bibinfo {author} {\bibfnamefont {S.}~\bibnamefont {Yang}}, \bibinfo
  {author} {\bibfnamefont {H.}~\bibnamefont {Li}}, \bibinfo {author}
  {\bibfnamefont {Y.}~\bibnamefont {Wu}}, \bibinfo {author} {\bibfnamefont
  {C.}~\bibnamefont {Liu}}, \bibinfo {author} {\bibfnamefont {J.}~\bibnamefont
  {Wang}}, \bibinfo {author} {\bibfnamefont {J.}~\bibnamefont {Shen}}, \bibinfo
  {author} {\bibfnamefont {J.}~\bibnamefont {Zhang}}, \bibinfo {author}
  {\bibfnamefont {Y.}~\bibnamefont {Wang}},\ and\ \bibinfo {author}
  {\bibfnamefont {Y.}~\bibnamefont {Wang}},\ }\bibfield  {title} {\bibinfo
  {title} {Direct observation of chiral edge current at zero magnetic field in
  a magnetic topological insulator},\ }\href
  {https://doi.org/10.1038/s41467-025-56326-7} {\bibfield  {journal} {\bibinfo
  {journal} {Nature Communications}\ }\textbf {\bibinfo {volume} {16}},\
  \bibinfo {pages} {963} (\bibinfo {year} {2025})}\BibitemShut {NoStop}%
\bibitem [{\citenamefont {Deng}\ \emph {et~al.}(2020)\citenamefont {Deng},
  \citenamefont {Yu}, \citenamefont {Shi}, \citenamefont {Guo}, \citenamefont
  {Xu}, \citenamefont {Wang}, \citenamefont {Chen},\ and\ \citenamefont
  {Zhang}}]{Deng2020}%
  \BibitemOpen
  \bibfield  {author} {\bibinfo {author} {\bibfnamefont {Y.}~\bibnamefont
  {Deng}}, \bibinfo {author} {\bibfnamefont {Y.}~\bibnamefont {Yu}}, \bibinfo
  {author} {\bibfnamefont {M.~Z.}\ \bibnamefont {Shi}}, \bibinfo {author}
  {\bibfnamefont {Z.}~\bibnamefont {Guo}}, \bibinfo {author} {\bibfnamefont
  {Z.}~\bibnamefont {Xu}}, \bibinfo {author} {\bibfnamefont {J.}~\bibnamefont
  {Wang}}, \bibinfo {author} {\bibfnamefont {X.~H.}\ \bibnamefont {Chen}},\
  and\ \bibinfo {author} {\bibfnamefont {Y.}~\bibnamefont {Zhang}},\ }\bibfield
   {title} {\bibinfo {title} {Quantum anomalous {H}all effect in intrinsic
  magnetic topological insulator {MnBi}$_2${Te}$_4$},\ }\href
  {https://doi.org/10.1126/science.aax8156} {\bibfield  {journal} {\bibinfo
  {journal} {Science}\ }\textbf {\bibinfo {volume} {367}},\ \bibinfo {pages}
  {895} (\bibinfo {year} {2020})}\BibitemShut {NoStop}%
\bibitem [{\citenamefont {Liu}\ \emph {et~al.}(2020)\citenamefont {Liu},
  \citenamefont {Wang}, \citenamefont {Li}, \citenamefont {Wu}, \citenamefont
  {Li}, \citenamefont {Li}, \citenamefont {He}, \citenamefont {Xu},
  \citenamefont {Zhang},\ and\ \citenamefont {Wang}}]{Liu2020}%
  \BibitemOpen
  \bibfield  {author} {\bibinfo {author} {\bibfnamefont {C.}~\bibnamefont
  {Liu}}, \bibinfo {author} {\bibfnamefont {Y.}~\bibnamefont {Wang}}, \bibinfo
  {author} {\bibfnamefont {H.}~\bibnamefont {Li}}, \bibinfo {author}
  {\bibfnamefont {Y.}~\bibnamefont {Wu}}, \bibinfo {author} {\bibfnamefont
  {Y.}~\bibnamefont {Li}}, \bibinfo {author} {\bibfnamefont {J.}~\bibnamefont
  {Li}}, \bibinfo {author} {\bibfnamefont {K.}~\bibnamefont {He}}, \bibinfo
  {author} {\bibfnamefont {Y.}~\bibnamefont {Xu}}, \bibinfo {author}
  {\bibfnamefont {J.}~\bibnamefont {Zhang}},\ and\ \bibinfo {author}
  {\bibfnamefont {Y.}~\bibnamefont {Wang}},\ }\bibfield  {title} {\bibinfo
  {title} {Robust axion insulator and {Chern} insulator phases in a
  two-dimensional antiferromagnetic topological insulator},\ }\href
  {https://doi.org/10.1038/s41563-019-0573-3} {\bibfield  {journal} {\bibinfo
  {journal} {Nature Materials}\ }\textbf {\bibinfo {volume} {19}},\ \bibinfo
  {pages} {522} (\bibinfo {year} {2020})}\BibitemShut {NoStop}%
\bibitem [{\citenamefont {Li}\ \emph {et~al.}(2021)\citenamefont {Li},
  \citenamefont {Jiang}, \citenamefont {Shen}, \citenamefont {Zhang},
  \citenamefont {Li}, \citenamefont {Tao}, \citenamefont {Devakul},
  \citenamefont {Watanabe}, \citenamefont {Taniguchi}, \citenamefont {Fu},
  \citenamefont {Shan},\ and\ \citenamefont {Mak}}]{Li2021}%
  \BibitemOpen
  \bibfield  {author} {\bibinfo {author} {\bibfnamefont {T.}~\bibnamefont
  {Li}}, \bibinfo {author} {\bibfnamefont {S.}~\bibnamefont {Jiang}}, \bibinfo
  {author} {\bibfnamefont {B.}~\bibnamefont {Shen}}, \bibinfo {author}
  {\bibfnamefont {Y.}~\bibnamefont {Zhang}}, \bibinfo {author} {\bibfnamefont
  {L.}~\bibnamefont {Li}}, \bibinfo {author} {\bibfnamefont {Z.}~\bibnamefont
  {Tao}}, \bibinfo {author} {\bibfnamefont {T.}~\bibnamefont {Devakul}},
  \bibinfo {author} {\bibfnamefont {K.}~\bibnamefont {Watanabe}}, \bibinfo
  {author} {\bibfnamefont {T.}~\bibnamefont {Taniguchi}}, \bibinfo {author}
  {\bibfnamefont {L.}~\bibnamefont {Fu}}, \bibinfo {author} {\bibfnamefont
  {J.}~\bibnamefont {Shan}},\ and\ \bibinfo {author} {\bibfnamefont {K.~F.}\
  \bibnamefont {Mak}},\ }\bibfield  {title} {\bibinfo {title} {Quantum
  anomalous {H}all effect from intertwined moiré bands},\ }\href
  {https://doi.org/10.1038/s41586-021-04171-1} {\bibfield  {journal} {\bibinfo
  {journal} {Nature}\ }\textbf {\bibinfo {volume} {600}},\ \bibinfo {pages}
  {641} (\bibinfo {year} {2021})}\BibitemShut {NoStop}%
\bibitem [{\citenamefont {Serlin}\ \emph {et~al.}(2020)\citenamefont {Serlin},
  \citenamefont {Tschirhart}, \citenamefont {Polshyn}, \citenamefont {Zhang},
  \citenamefont {Zhu}, \citenamefont {Watanabe}, \citenamefont {Taniguchi},
  \citenamefont {Balents},\ and\ \citenamefont {Young}}]{Serlin2020}%
  \BibitemOpen
  \bibfield  {author} {\bibinfo {author} {\bibfnamefont {M.}~\bibnamefont
  {Serlin}}, \bibinfo {author} {\bibfnamefont {C.~L.}\ \bibnamefont
  {Tschirhart}}, \bibinfo {author} {\bibfnamefont {H.}~\bibnamefont {Polshyn}},
  \bibinfo {author} {\bibfnamefont {Y.}~\bibnamefont {Zhang}}, \bibinfo
  {author} {\bibfnamefont {J.}~\bibnamefont {Zhu}}, \bibinfo {author}
  {\bibfnamefont {K.}~\bibnamefont {Watanabe}}, \bibinfo {author}
  {\bibfnamefont {T.}~\bibnamefont {Taniguchi}}, \bibinfo {author}
  {\bibfnamefont {L.}~\bibnamefont {Balents}},\ and\ \bibinfo {author}
  {\bibfnamefont {A.~F.}\ \bibnamefont {Young}},\ }\bibfield  {title} {\bibinfo
  {title} {Intrinsic quantized anomalous {H}all effect in a moiré
  heterostructure},\ }\href {https://doi.org/10.1126/science.aay5533}
  {\bibfield  {journal} {\bibinfo  {journal} {Science}\ }\textbf {\bibinfo
  {volume} {367}},\ \bibinfo {pages} {900} (\bibinfo {year}
  {2020})}\BibitemShut {NoStop}%
\bibitem [{\citenamefont {Baltz}\ \emph {et~al.}(2018)\citenamefont {Baltz},
  \citenamefont {Manchon}, \citenamefont {Tsoi}, \citenamefont {Moriyama},
  \citenamefont {Ono},\ and\ \citenamefont {Tserkovnyak}}]{Baltz2018}%
  \BibitemOpen
  \bibfield  {author} {\bibinfo {author} {\bibfnamefont {V.}~\bibnamefont
  {Baltz}}, \bibinfo {author} {\bibfnamefont {A.}~\bibnamefont {Manchon}},
  \bibinfo {author} {\bibfnamefont {M.}~\bibnamefont {Tsoi}}, \bibinfo {author}
  {\bibfnamefont {T.}~\bibnamefont {Moriyama}}, \bibinfo {author}
  {\bibfnamefont {T.}~\bibnamefont {Ono}},\ and\ \bibinfo {author}
  {\bibfnamefont {Y.}~\bibnamefont {Tserkovnyak}},\ }\bibfield  {title}
  {\bibinfo {title} {Antiferromagnetic spintronics},\ }\href
  {https://doi.org/10.1103/RevModPhys.90.015005} {\bibfield  {journal}
  {\bibinfo  {journal} {Rev. Mod. Phys.}\ }\textbf {\bibinfo {volume} {90}},\
  \bibinfo {pages} {015005} (\bibinfo {year} {2018})}\BibitemShut {NoStop}%
\bibitem [{\citenamefont {Hafez-Torbati}\ \emph {et~al.}(2021)\citenamefont
  {Hafez-Torbati}, \citenamefont {Bossini}, \citenamefont {Anders},\ and\
  \citenamefont {Uhrig}}]{Hafez-Torbati2021}%
  \BibitemOpen
  \bibfield  {author} {\bibinfo {author} {\bibfnamefont {M.}~\bibnamefont
  {Hafez-Torbati}}, \bibinfo {author} {\bibfnamefont {D.}~\bibnamefont
  {Bossini}}, \bibinfo {author} {\bibfnamefont {F.~B.}\ \bibnamefont
  {Anders}},\ and\ \bibinfo {author} {\bibfnamefont {G.~S.}\ \bibnamefont
  {Uhrig}},\ }\bibfield  {title} {\bibinfo {title} {Magnetic blue shift of
  {Mott} gaps enhanced by double exchange},\ }\href
  {https://doi.org/10.1103/PhysRevResearch.3.043232} {\bibfield  {journal}
  {\bibinfo  {journal} {Phys. Rev. Research}\ }\textbf {\bibinfo {volume}
  {3}},\ \bibinfo {pages} {043232} (\bibinfo {year} {2021})}\BibitemShut
  {NoStop}%
\bibitem [{\citenamefont {Hafez-Torbati}\ \emph {et~al.}(2022)\citenamefont
  {Hafez-Torbati}, \citenamefont {Anders},\ and\ \citenamefont
  {Uhrig}}]{Hafez-Torbati2022}%
  \BibitemOpen
  \bibfield  {author} {\bibinfo {author} {\bibfnamefont {M.}~\bibnamefont
  {Hafez-Torbati}}, \bibinfo {author} {\bibfnamefont {F.~B.}\ \bibnamefont
  {Anders}},\ and\ \bibinfo {author} {\bibfnamefont {G.~S.}\ \bibnamefont
  {Uhrig}},\ }\bibfield  {title} {\bibinfo {title} {Simplified approach to the
  magnetic blue shift of {Mott} gaps},\ }\href
  {https://doi.org/10.1103/PhysRevB.106.205117} {\bibfield  {journal} {\bibinfo
   {journal} {Phys. Rev. B}\ }\textbf {\bibinfo {volume} {106}},\ \bibinfo
  {pages} {205117} (\bibinfo {year} {2022})}\BibitemShut {NoStop}%
\bibitem [{\citenamefont {Bossini}\ \emph {et~al.}(2020)\citenamefont
  {Bossini}, \citenamefont {Terschanski}, \citenamefont {Mertens},
  \citenamefont {Springholz}, \citenamefont {Bonanni}, \citenamefont {Uhrig},\
  and\ \citenamefont {Cinchetti}}]{Bossini2020}%
  \BibitemOpen
  \bibfield  {author} {\bibinfo {author} {\bibfnamefont {D.}~\bibnamefont
  {Bossini}}, \bibinfo {author} {\bibfnamefont {M.}~\bibnamefont
  {Terschanski}}, \bibinfo {author} {\bibfnamefont {F.}~\bibnamefont
  {Mertens}}, \bibinfo {author} {\bibfnamefont {G.}~\bibnamefont {Springholz}},
  \bibinfo {author} {\bibfnamefont {A.}~\bibnamefont {Bonanni}}, \bibinfo
  {author} {\bibfnamefont {G.~S.}\ \bibnamefont {Uhrig}},\ and\ \bibinfo
  {author} {\bibfnamefont {M.}~\bibnamefont {Cinchetti}},\ }\bibfield  {title}
  {\bibinfo {title} {Exchange-mediated magnetic blue-shift of the band-gap
  energy in the antiferromagnetic semiconductor {MnTe}},\ }\href
  {https://doi.org/10.1088/1367-2630/aba0e7} {\bibfield  {journal} {\bibinfo
  {journal} {New Journal of Physics}\ }\textbf {\bibinfo {volume} {22}},\
  \bibinfo {pages} {083029} (\bibinfo {year} {2020})}\BibitemShut {NoStop}%
\bibitem [{\citenamefont {Sangiovanni}\ \emph {et~al.}(2006)\citenamefont
  {Sangiovanni}, \citenamefont {Toschi}, \citenamefont {Koch}, \citenamefont
  {Held}, \citenamefont {Capone}, \citenamefont {Castellani}, \citenamefont
  {Gunnarsson}, \citenamefont {Mo}, \citenamefont {Allen}, \citenamefont {Kim},
  \citenamefont {Sekiyama}, \citenamefont {Yamasaki}, \citenamefont {Suga},\
  and\ \citenamefont {Metcalf}}]{Sangiovanni2006}%
  \BibitemOpen
  \bibfield  {author} {\bibinfo {author} {\bibfnamefont {G.}~\bibnamefont
  {Sangiovanni}}, \bibinfo {author} {\bibfnamefont {A.}~\bibnamefont {Toschi}},
  \bibinfo {author} {\bibfnamefont {E.}~\bibnamefont {Koch}}, \bibinfo {author}
  {\bibfnamefont {K.}~\bibnamefont {Held}}, \bibinfo {author} {\bibfnamefont
  {M.}~\bibnamefont {Capone}}, \bibinfo {author} {\bibfnamefont
  {C.}~\bibnamefont {Castellani}}, \bibinfo {author} {\bibfnamefont
  {O.}~\bibnamefont {Gunnarsson}}, \bibinfo {author} {\bibfnamefont {S.-K.}\
  \bibnamefont {Mo}}, \bibinfo {author} {\bibfnamefont {J.~W.}\ \bibnamefont
  {Allen}}, \bibinfo {author} {\bibfnamefont {H.-D.}\ \bibnamefont {Kim}},
  \bibinfo {author} {\bibfnamefont {A.}~\bibnamefont {Sekiyama}}, \bibinfo
  {author} {\bibfnamefont {A.}~\bibnamefont {Yamasaki}}, \bibinfo {author}
  {\bibfnamefont {S.}~\bibnamefont {Suga}},\ and\ \bibinfo {author}
  {\bibfnamefont {P.}~\bibnamefont {Metcalf}},\ }\bibfield  {title} {\bibinfo
  {title} {Static versus dynamical mean-field theory of {Mott}
  antiferromagnets},\ }\href {https://doi.org/10.1103/PhysRevB.73.205121}
  {\bibfield  {journal} {\bibinfo  {journal} {Phys. Rev. B}\ }\textbf {\bibinfo
  {volume} {73}},\ \bibinfo {pages} {205121} (\bibinfo {year}
  {2006})}\BibitemShut {NoStop}%
\bibitem [{\citenamefont {Hao}\ \emph {et~al.}(2019)\citenamefont {Hao},
  \citenamefont {Liu}, \citenamefont {Feng}, \citenamefont {Ma}, \citenamefont
  {Schwier}, \citenamefont {Arita}, \citenamefont {Kumar}, \citenamefont {Hu},
  \citenamefont {Lu}, \citenamefont {Zeng}, \citenamefont {Wang}, \citenamefont
  {Hao}, \citenamefont {Sun}, \citenamefont {Zhang}, \citenamefont {Mei},
  \citenamefont {Ni}, \citenamefont {Wu}, \citenamefont {Shimada},
  \citenamefont {Chen}, \citenamefont {Liu},\ and\ \citenamefont
  {Liu}}]{Hao2019}%
  \BibitemOpen
  \bibfield  {author} {\bibinfo {author} {\bibfnamefont {Y.-J.}\ \bibnamefont
  {Hao}}, \bibinfo {author} {\bibfnamefont {P.}~\bibnamefont {Liu}}, \bibinfo
  {author} {\bibfnamefont {Y.}~\bibnamefont {Feng}}, \bibinfo {author}
  {\bibfnamefont {X.-M.}\ \bibnamefont {Ma}}, \bibinfo {author} {\bibfnamefont
  {E.~F.}\ \bibnamefont {Schwier}}, \bibinfo {author} {\bibfnamefont
  {M.}~\bibnamefont {Arita}}, \bibinfo {author} {\bibfnamefont
  {S.}~\bibnamefont {Kumar}}, \bibinfo {author} {\bibfnamefont
  {C.}~\bibnamefont {Hu}}, \bibinfo {author} {\bibfnamefont {R.}~\bibnamefont
  {Lu}}, \bibinfo {author} {\bibfnamefont {M.}~\bibnamefont {Zeng}}, \bibinfo
  {author} {\bibfnamefont {Y.}~\bibnamefont {Wang}}, \bibinfo {author}
  {\bibfnamefont {Z.}~\bibnamefont {Hao}}, \bibinfo {author} {\bibfnamefont
  {H.-Y.}\ \bibnamefont {Sun}}, \bibinfo {author} {\bibfnamefont
  {K.}~\bibnamefont {Zhang}}, \bibinfo {author} {\bibfnamefont
  {J.}~\bibnamefont {Mei}}, \bibinfo {author} {\bibfnamefont {N.}~\bibnamefont
  {Ni}}, \bibinfo {author} {\bibfnamefont {L.}~\bibnamefont {Wu}}, \bibinfo
  {author} {\bibfnamefont {K.}~\bibnamefont {Shimada}}, \bibinfo {author}
  {\bibfnamefont {C.}~\bibnamefont {Chen}}, \bibinfo {author} {\bibfnamefont
  {Q.}~\bibnamefont {Liu}},\ and\ \bibinfo {author} {\bibfnamefont
  {C.}~\bibnamefont {Liu}},\ }\bibfield  {title} {\bibinfo {title} {Gapless
  {Surface Dirac Cone in Antiferromagnetic Topological Insulator
  {MnBi}$_2${Te}$_4$}},\ }\href {https://doi.org/10.1103/PhysRevX.9.041038}
  {\bibfield  {journal} {\bibinfo  {journal} {Phys. Rev. X}\ }\textbf {\bibinfo
  {volume} {9}},\ \bibinfo {pages} {041038} (\bibinfo {year}
  {2019})}\BibitemShut {NoStop}%
\bibitem [{\citenamefont {Chen}\ \emph {et~al.}(2019)\citenamefont {Chen},
  \citenamefont {Xu}, \citenamefont {Li}, \citenamefont {Li}, \citenamefont
  {Wang}, \citenamefont {Zhang}, \citenamefont {Li}, \citenamefont {Wu},
  \citenamefont {Liang}, \citenamefont {Chen}, \citenamefont {Jung},
  \citenamefont {Cacho}, \citenamefont {Mao}, \citenamefont {Liu},
  \citenamefont {Wang}, \citenamefont {Guo}, \citenamefont {Xu}, \citenamefont
  {Liu}, \citenamefont {Yang},\ and\ \citenamefont {Chen}}]{Chen2019}%
  \BibitemOpen
  \bibfield  {author} {\bibinfo {author} {\bibfnamefont {Y.~J.}\ \bibnamefont
  {Chen}}, \bibinfo {author} {\bibfnamefont {L.~X.}\ \bibnamefont {Xu}},
  \bibinfo {author} {\bibfnamefont {J.~H.}\ \bibnamefont {Li}}, \bibinfo
  {author} {\bibfnamefont {Y.~W.}\ \bibnamefont {Li}}, \bibinfo {author}
  {\bibfnamefont {H.~Y.}\ \bibnamefont {Wang}}, \bibinfo {author}
  {\bibfnamefont {C.~F.}\ \bibnamefont {Zhang}}, \bibinfo {author}
  {\bibfnamefont {H.}~\bibnamefont {Li}}, \bibinfo {author} {\bibfnamefont
  {Y.}~\bibnamefont {Wu}}, \bibinfo {author} {\bibfnamefont {A.~J.}\
  \bibnamefont {Liang}}, \bibinfo {author} {\bibfnamefont {C.}~\bibnamefont
  {Chen}}, \bibinfo {author} {\bibfnamefont {S.~W.}\ \bibnamefont {Jung}},
  \bibinfo {author} {\bibfnamefont {C.}~\bibnamefont {Cacho}}, \bibinfo
  {author} {\bibfnamefont {Y.~H.}\ \bibnamefont {Mao}}, \bibinfo {author}
  {\bibfnamefont {S.}~\bibnamefont {Liu}}, \bibinfo {author} {\bibfnamefont
  {M.~X.}\ \bibnamefont {Wang}}, \bibinfo {author} {\bibfnamefont {Y.~F.}\
  \bibnamefont {Guo}}, \bibinfo {author} {\bibfnamefont {Y.}~\bibnamefont
  {Xu}}, \bibinfo {author} {\bibfnamefont {Z.~K.}\ \bibnamefont {Liu}},
  \bibinfo {author} {\bibfnamefont {L.~X.}\ \bibnamefont {Yang}},\ and\
  \bibinfo {author} {\bibfnamefont {Y.~L.}\ \bibnamefont {Chen}},\ }\bibfield
  {title} {\bibinfo {title} {Topological {Electronic Structure and Its
  Temperature Evolution in Antiferromagnetic Topological Insulator
  {MnBi}$_2${Te}$_4$}},\ }\href {https://doi.org/10.1103/PhysRevX.9.041040}
  {\bibfield  {journal} {\bibinfo  {journal} {Phys. Rev. X}\ }\textbf {\bibinfo
  {volume} {9}},\ \bibinfo {pages} {041040} (\bibinfo {year}
  {2019})}\BibitemShut {NoStop}%
\bibitem [{\citenamefont {Li}\ \emph {et~al.}(2019)\citenamefont {Li},
  \citenamefont {Gao}, \citenamefont {Duan}, \citenamefont {Xu}, \citenamefont
  {Zhu}, \citenamefont {Tian}, \citenamefont {Gao}, \citenamefont {Fan},
  \citenamefont {Rao}, \citenamefont {Huang}, \citenamefont {Li}, \citenamefont
  {Yan}, \citenamefont {Liu}, \citenamefont {Liu}, \citenamefont {Huang},
  \citenamefont {Li}, \citenamefont {Liu}, \citenamefont {Zhang}, \citenamefont
  {Zhang}, \citenamefont {Kondo}, \citenamefont {Shin}, \citenamefont {Lei},
  \citenamefont {Shi}, \citenamefont {Zhang}, \citenamefont {Weng},
  \citenamefont {Qian},\ and\ \citenamefont {Ding}}]{Li2019b}%
  \BibitemOpen
  \bibfield  {author} {\bibinfo {author} {\bibfnamefont {H.}~\bibnamefont
  {Li}}, \bibinfo {author} {\bibfnamefont {S.-Y.}\ \bibnamefont {Gao}},
  \bibinfo {author} {\bibfnamefont {S.-F.}\ \bibnamefont {Duan}}, \bibinfo
  {author} {\bibfnamefont {Y.-F.}\ \bibnamefont {Xu}}, \bibinfo {author}
  {\bibfnamefont {K.-J.}\ \bibnamefont {Zhu}}, \bibinfo {author} {\bibfnamefont
  {S.-J.}\ \bibnamefont {Tian}}, \bibinfo {author} {\bibfnamefont {J.-C.}\
  \bibnamefont {Gao}}, \bibinfo {author} {\bibfnamefont {W.-H.}\ \bibnamefont
  {Fan}}, \bibinfo {author} {\bibfnamefont {Z.-C.}\ \bibnamefont {Rao}},
  \bibinfo {author} {\bibfnamefont {J.-R.}\ \bibnamefont {Huang}}, \bibinfo
  {author} {\bibfnamefont {J.-J.}\ \bibnamefont {Li}}, \bibinfo {author}
  {\bibfnamefont {D.-Y.}\ \bibnamefont {Yan}}, \bibinfo {author} {\bibfnamefont
  {Z.-T.}\ \bibnamefont {Liu}}, \bibinfo {author} {\bibfnamefont {W.-L.}\
  \bibnamefont {Liu}}, \bibinfo {author} {\bibfnamefont {Y.-B.}\ \bibnamefont
  {Huang}}, \bibinfo {author} {\bibfnamefont {Y.-L.}\ \bibnamefont {Li}},
  \bibinfo {author} {\bibfnamefont {Y.}~\bibnamefont {Liu}}, \bibinfo {author}
  {\bibfnamefont {G.-B.}\ \bibnamefont {Zhang}}, \bibinfo {author}
  {\bibfnamefont {P.}~\bibnamefont {Zhang}}, \bibinfo {author} {\bibfnamefont
  {T.}~\bibnamefont {Kondo}}, \bibinfo {author} {\bibfnamefont
  {S.}~\bibnamefont {Shin}}, \bibinfo {author} {\bibfnamefont {H.-C.}\
  \bibnamefont {Lei}}, \bibinfo {author} {\bibfnamefont {Y.-G.}\ \bibnamefont
  {Shi}}, \bibinfo {author} {\bibfnamefont {W.-T.}\ \bibnamefont {Zhang}},
  \bibinfo {author} {\bibfnamefont {H.-M.}\ \bibnamefont {Weng}}, \bibinfo
  {author} {\bibfnamefont {T.}~\bibnamefont {Qian}},\ and\ \bibinfo {author}
  {\bibfnamefont {H.}~\bibnamefont {Ding}},\ }\bibfield  {title} {\bibinfo
  {title} {{Dirac Surface States in Intrinsic Magnetic Topological Insulators
  {EuSn}$_2${As}$_2$ and {MnBi}$_{2n}${Te}$_{3n+1}$}},\ }\href
  {https://doi.org/10.1103/PhysRevX.9.041039} {\bibfield  {journal} {\bibinfo
  {journal} {Physical Review X}\ }\textbf {\bibinfo {volume} {9}},\ \bibinfo
  {pages} {041039} (\bibinfo {year} {2019})}\BibitemShut {NoStop}%
\bibitem [{\citenamefont {Swatek}\ \emph {et~al.}(2020)\citenamefont {Swatek},
  \citenamefont {Wu}, \citenamefont {Wang}, \citenamefont {Lee}, \citenamefont
  {Schrunk}, \citenamefont {Yan},\ and\ \citenamefont {Kaminski}}]{Swatek2020}%
  \BibitemOpen
  \bibfield  {author} {\bibinfo {author} {\bibfnamefont {P.}~\bibnamefont
  {Swatek}}, \bibinfo {author} {\bibfnamefont {Y.}~\bibnamefont {Wu}}, \bibinfo
  {author} {\bibfnamefont {L.-L.}\ \bibnamefont {Wang}}, \bibinfo {author}
  {\bibfnamefont {K.}~\bibnamefont {Lee}}, \bibinfo {author} {\bibfnamefont
  {B.}~\bibnamefont {Schrunk}}, \bibinfo {author} {\bibfnamefont
  {J.}~\bibnamefont {Yan}},\ and\ \bibinfo {author} {\bibfnamefont
  {A.}~\bibnamefont {Kaminski}},\ }\bibfield  {title} {\bibinfo {title}
  {Gapless {D}irac surface states in the antiferromagnetic topological
  insulator {MnBi}$_2${Te}$_4$},\ }\href
  {https://doi.org/10.1103/PhysRevB.101.161109} {\bibfield  {journal} {\bibinfo
   {journal} {Phys. Rev. B}\ }\textbf {\bibinfo {volume} {101}},\ \bibinfo
  {pages} {161109} (\bibinfo {year} {2020})}\BibitemShut {NoStop}%
\bibitem [{\citenamefont {Jiang}\ \emph {et~al.}(2018)\citenamefont {Jiang},
  \citenamefont {Zhou}, \citenamefont {Dai},\ and\ \citenamefont
  {Wang}}]{Jiang2018}%
  \BibitemOpen
  \bibfield  {author} {\bibinfo {author} {\bibfnamefont {K.}~\bibnamefont
  {Jiang}}, \bibinfo {author} {\bibfnamefont {S.}~\bibnamefont {Zhou}},
  \bibinfo {author} {\bibfnamefont {X.}~\bibnamefont {Dai}},\ and\ \bibinfo
  {author} {\bibfnamefont {Z.}~\bibnamefont {Wang}},\ }\bibfield  {title}
  {\bibinfo {title} {Antiferromagnetic {Chern Insulators in Noncentrosymmetric
  Systems}},\ }\href {https://link.aps.org/doi/10.1103/PhysRevLett.120.157205}
  {\bibfield  {journal} {\bibinfo  {journal} {Phys. Rev. Lett.}\ }\textbf
  {\bibinfo {volume} {120}},\ \bibinfo {pages} {157205} (\bibinfo {year}
  {2018})}\BibitemShut {NoStop}%
\bibitem [{\citenamefont {Ebrahimkhas}\ \emph {et~al.}(2022)\citenamefont
  {Ebrahimkhas}, \citenamefont {Uhrig}, \citenamefont {Hofstetter},\ and\
  \citenamefont {Hafez-Torbati}}]{Ebrahimkhas2022}%
  \BibitemOpen
  \bibfield  {author} {\bibinfo {author} {\bibfnamefont {M.}~\bibnamefont
  {Ebrahimkhas}}, \bibinfo {author} {\bibfnamefont {G.~S.}\ \bibnamefont
  {Uhrig}}, \bibinfo {author} {\bibfnamefont {W.}~\bibnamefont {Hofstetter}},\
  and\ \bibinfo {author} {\bibfnamefont {M.}~\bibnamefont {Hafez-Torbati}},\
  }\bibfield  {title} {\bibinfo {title} {Antiferromagnetic {C}hern insulator in
  centrosymmetric systems},\ }\href
  {https://doi.org/10.1103/PhysRevB.106.205107} {\bibfield  {journal} {\bibinfo
   {journal} {Phys. Rev. B}\ }\textbf {\bibinfo {volume} {106}},\ \bibinfo
  {pages} {205107} (\bibinfo {year} {2022})}\BibitemShut {NoStop}%
\bibitem [{\citenamefont {Guo}\ \emph {et~al.}(2023)\citenamefont {Guo},
  \citenamefont {Liu},\ and\ \citenamefont {Lu}}]{Guo2023}%
  \BibitemOpen
  \bibfield  {author} {\bibinfo {author} {\bibfnamefont {P.-J.}\ \bibnamefont
  {Guo}}, \bibinfo {author} {\bibfnamefont {Z.-X.}\ \bibnamefont {Liu}},\ and\
  \bibinfo {author} {\bibfnamefont {Z.-Y.}\ \bibnamefont {Lu}},\ }\bibfield
  {title} {\bibinfo {title} {Quantum anomalous {H}all effect in collinear
  antiferromagnetism},\ }\href {https://doi.org/10.1038/s41524-023-01025-4}
  {\bibfield  {journal} {\bibinfo  {journal} {npj Computational Materials}\
  }\textbf {\bibinfo {volume} {9}},\ \bibinfo {pages} {70} (\bibinfo {year}
  {2023})}\BibitemShut {NoStop}%
\bibitem [{\citenamefont {Hafez-Torbati}(2024)}]{Hafez-Torbati2024a}%
  \BibitemOpen
  \bibfield  {author} {\bibinfo {author} {\bibfnamefont {M.}~\bibnamefont
  {Hafez-Torbati}},\ }\bibfield  {title} {\bibinfo {title} {Antiferromagnetic
  topological insulators in heavy-fermion systems},\ }\href
  {https://doi.org/10.1103/PhysRevB.110.115147} {\bibfield  {journal} {\bibinfo
   {journal} {Phys. Rev. B}\ }\textbf {\bibinfo {volume} {110}},\ \bibinfo
  {pages} {115147} (\bibinfo {year} {2024})}\BibitemShut {NoStop}%
\bibitem [{\citenamefont {Hafez-Torbati}\ and\ \citenamefont
  {Uhrig}(2024)}]{Hafez-Torbati2024}%
  \BibitemOpen
  \bibfield  {author} {\bibinfo {author} {\bibfnamefont {M.}~\bibnamefont
  {Hafez-Torbati}}\ and\ \bibinfo {author} {\bibfnamefont {G.~S.}\ \bibnamefont
  {Uhrig}},\ }\bibfield  {title} {\bibinfo {title} {Antiferromagnetic Chern
  insulator with large charge gap in heavy transition-metal compounds},\ }\href
  {https://doi.org/10.1038/s41598-024-68044-z} {\bibfield  {journal} {\bibinfo
  {journal} {Scientific Reports}\ }\textbf {\bibinfo {volume} {14}},\ \bibinfo
  {pages} {17168} (\bibinfo {year} {2024})}\BibitemShut {NoStop}%
\bibitem [{\citenamefont {Wu}\ \emph {et~al.}(2023)\citenamefont {Wu},
  \citenamefont {Song}, \citenamefont {Ji}, \citenamefont {Wang}, \citenamefont
  {Zhang},\ and\ \citenamefont {Zhang}}]{Wu2023}%
  \BibitemOpen
  \bibfield  {author} {\bibinfo {author} {\bibfnamefont {B.}~\bibnamefont
  {Wu}}, \bibinfo {author} {\bibfnamefont {Y.-l.}\ \bibnamefont {Song}},
  \bibinfo {author} {\bibfnamefont {W.-x.}\ \bibnamefont {Ji}}, \bibinfo
  {author} {\bibfnamefont {P.-j.}\ \bibnamefont {Wang}}, \bibinfo {author}
  {\bibfnamefont {S.-f.}\ \bibnamefont {Zhang}},\ and\ \bibinfo {author}
  {\bibfnamefont {C.-w.}\ \bibnamefont {Zhang}},\ }\bibfield  {title} {\bibinfo
  {title} {Quantum anomalous {Hall effect in an antiferromagnetic monolayer of
  MoO}},\ }\href {https://doi.org/10.1103/PhysRevB.107.214419} {\bibfield
  {journal} {\bibinfo  {journal} {Phys. Rev. B}\ }\textbf {\bibinfo {volume}
  {107}},\ \bibinfo {pages} {214419} (\bibinfo {year} {2023})}\BibitemShut
  {NoStop}%
\bibitem [{\citenamefont {Fazekas}(1999)}]{Fazekas1999}%
  \BibitemOpen
  \bibfield  {author} {\bibinfo {author} {\bibfnamefont {P.}~\bibnamefont
  {Fazekas}},\ }\href {http://books.google.de/books?id=j2F1uQAACAAJ} {\emph
  {\bibinfo {title} {Lecture Notes on Electron Correlation and Magnetism}}},\
  Series in modern condensed matter physics\ (\bibinfo  {publisher} {World
  Scientific},\ \bibinfo {year} {1999})\BibitemShut {NoStop}%
\bibitem [{sm()}]{sm}%
  \BibitemOpen
  \href@noop {} {\bibinfo  {journal} {See Supplemental Materials, which
  includes Refs. \cite{Wang2012,Hawashin2024}, for the DMFT solution of Eq.
  (2), for a discussion of the ionicity, for the analysis of the Hall
  conductance in Eq. (3), for the mean-field N\'eel temperature of the
  effective spin model, for the calculation of the spectral function on the
  cylindrical geometry, and for further details of the phase diagram in Fig.
  2}\ }\BibitemShut {NoStop}%
\bibitem [{\citenamefont {Wang}\ and\ \citenamefont {Zhang}(2012)}]{Wang2012}%
  \BibitemOpen
  \bibfield  {author} {\bibinfo {author} {\bibfnamefont {Z.}~\bibnamefont
  {Wang}}\ and\ \bibinfo {author} {\bibfnamefont {S.-C.}\ \bibnamefont
  {Zhang}},\ }\bibfield  {title} {\bibinfo {title} {Simplified {Topological
  Invariants for Interacting Insulators}},\ }\href
  {https://doi.org/10.1103/PhysRevX.2.031008} {\bibfield  {journal} {\bibinfo
  {journal} {Phys. Rev. X}\ }\textbf {\bibinfo {volume} {2}},\ \bibinfo {pages}
  {031008} (\bibinfo {year} {2012})}\BibitemShut {NoStop}%
\bibitem [{\citenamefont {Hawashin}\ \emph {et~al.}(2024)\citenamefont
  {Hawashin}, \citenamefont {Sirker},\ and\ \citenamefont
  {Uhrig}}]{Hawashin2024}%
  \BibitemOpen
  \bibfield  {author} {\bibinfo {author} {\bibfnamefont {B.}~\bibnamefont
  {Hawashin}}, \bibinfo {author} {\bibfnamefont {J.}~\bibnamefont {Sirker}},\
  and\ \bibinfo {author} {\bibfnamefont {G.~S.}\ \bibnamefont {Uhrig}},\
  }\bibfield  {title} {\bibinfo {title} {Topological properties of
  single-particle states decaying into a continuum due to interaction},\ }\href
  {https://doi.org/10.1103/PhysRevResearch.6.L042041} {\bibfield  {journal}
  {\bibinfo  {journal} {Phys. Rev. Res.}\ }\textbf {\bibinfo {volume} {6}},\
  \bibinfo {pages} {L042041} (\bibinfo {year} {2024})}\BibitemShut {NoStop}%
\bibitem [{\citenamefont {Kane}\ and\ \citenamefont {Mele}(2005)}]{Kane2005b}%
  \BibitemOpen
\bibfield  {journal} {  }\bibfield  {author} {\bibinfo {author} {\bibfnamefont
  {C.~L.}\ \bibnamefont {Kane}}\ and\ \bibinfo {author} {\bibfnamefont {E.~J.}\
  \bibnamefont {Mele}},\ }\bibfield  {title} {\bibinfo {title} {Quantum {Spin
  {H}all Effect in Graphene}},\ }\href
  {https://doi.org/10.1103/PhysRevLett.95.226801} {\bibfield  {journal}
  {\bibinfo  {journal} {Phys. Rev. Lett.}\ }\textbf {\bibinfo {volume} {95}},\
  \bibinfo {pages} {226801} (\bibinfo {year} {2005})}\BibitemShut {NoStop}%
\bibitem [{\citenamefont {Georges}\ \emph {et~al.}(1996)\citenamefont
  {Georges}, \citenamefont {Kotliar}, \citenamefont {Krauth},\ and\
  \citenamefont {Rozenberg}}]{Georges1996}%
  \BibitemOpen
  \bibfield  {author} {\bibinfo {author} {\bibfnamefont {A.}~\bibnamefont
  {Georges}}, \bibinfo {author} {\bibfnamefont {G.}~\bibnamefont {Kotliar}},
  \bibinfo {author} {\bibfnamefont {W.}~\bibnamefont {Krauth}},\ and\ \bibinfo
  {author} {\bibfnamefont {M.~J.}\ \bibnamefont {Rozenberg}},\ }\bibfield
  {title} {\bibinfo {title} {Dynamical mean-field theory of strongly correlated
  fermion systems and the limit of infinite dimensions},\ }\href
  {https://doi.org/10.1103/RevModPhys.68.13} {\bibfield  {journal} {\bibinfo
  {journal} {Rev. Mod. Phys.}\ }\textbf {\bibinfo {volume} {68}},\ \bibinfo
  {pages} {13} (\bibinfo {year} {1996})}\BibitemShut {NoStop}%
\bibitem [{\citenamefont {Caffarel}\ and\ \citenamefont
  {Krauth}(1994)}]{Caffarel1994}%
  \BibitemOpen
  \bibfield  {author} {\bibinfo {author} {\bibfnamefont {M.}~\bibnamefont
  {Caffarel}}\ and\ \bibinfo {author} {\bibfnamefont {W.}~\bibnamefont
  {Krauth}},\ }\bibfield  {title} {\bibinfo {title} {Exact diagonalization
  approach to correlated fermions in infinite dimensions: Mott transition and
  superconductivity},\ }\href
  {https://link.aps.org/doi/10.1103/PhysRevLett.72.1545} {\bibfield  {journal}
  {\bibinfo  {journal} {Phys. Rev. Lett.}\ }\textbf {\bibinfo {volume} {72}},\
  \bibinfo {pages} {1545} (\bibinfo {year} {1994})}\BibitemShut {NoStop}%
\bibitem [{\citenamefont {Potthoff}\ and\ \citenamefont
  {Nolting}(1999)}]{Potthoff1999}%
  \BibitemOpen
  \bibfield  {author} {\bibinfo {author} {\bibfnamefont {M.}~\bibnamefont
  {Potthoff}}\ and\ \bibinfo {author} {\bibfnamefont {W.}~\bibnamefont
  {Nolting}},\ }\bibfield  {title} {\bibinfo {title} {Surface metal-insulator
  transition in the {H}ubbard model},\ }\href
  {https://doi.org/10.1103/PhysRevB.59.2549} {\bibfield  {journal} {\bibinfo
  {journal} {Phys. Rev. B}\ }\textbf {\bibinfo {volume} {59}},\ \bibinfo
  {pages} {2549} (\bibinfo {year} {1999})}\BibitemShut {NoStop}%
\bibitem [{\citenamefont {Song}\ \emph {et~al.}(2008)\citenamefont {Song},
  \citenamefont {Wortis},\ and\ \citenamefont {Atkinson}}]{Song2008}%
  \BibitemOpen
  \bibfield  {author} {\bibinfo {author} {\bibfnamefont {Y.}~\bibnamefont
  {Song}}, \bibinfo {author} {\bibfnamefont {R.}~\bibnamefont {Wortis}},\ and\
  \bibinfo {author} {\bibfnamefont {W.~A.}\ \bibnamefont {Atkinson}},\
  }\bibfield  {title} {\bibinfo {title} {Dynamical mean field study of the
  two-dimensional disordered {H}ubbard model},\ }\href
  {https://doi.org/10.1103/PhysRevB.77.054202} {\bibfield  {journal} {\bibinfo
  {journal} {Phys. Rev. B}\ }\textbf {\bibinfo {volume} {77}},\ \bibinfo
  {pages} {054202} (\bibinfo {year} {2008})}\BibitemShut {NoStop}%
\bibitem [{\citenamefont {Snoek}\ \emph {et~al.}(2008)\citenamefont {Snoek},
  \citenamefont {Titvinidze}, \citenamefont {T{\H{o}}ke}, \citenamefont
  {Byczuk},\ and\ \citenamefont {Hofstetter}}]{Snoek2008}%
  \BibitemOpen
  \bibfield  {author} {\bibinfo {author} {\bibfnamefont {M.}~\bibnamefont
  {Snoek}}, \bibinfo {author} {\bibfnamefont {I.}~\bibnamefont {Titvinidze}},
  \bibinfo {author} {\bibfnamefont {C.}~\bibnamefont {T{\H{o}}ke}}, \bibinfo
  {author} {\bibfnamefont {K.}~\bibnamefont {Byczuk}},\ and\ \bibinfo {author}
  {\bibfnamefont {W.}~\bibnamefont {Hofstetter}},\ }\bibfield  {title}
  {\bibinfo {title} {Antiferromagnetic order of strongly interacting fermions
  in a trap: real-space dynamical mean-field analysis},\ }\href
  {https://doi.org/10.1088/1367-2630/10/9/093008} {\bibfield  {journal}
  {\bibinfo  {journal} {New Journal of Physics}\ }\textbf {\bibinfo {volume}
  {10}},\ \bibinfo {pages} {093008} (\bibinfo {year} {2008})}\BibitemShut
  {NoStop}%
\bibitem [{\citenamefont {Hafez-Torbati}\ and\ \citenamefont
  {Hofstetter}(2018)}]{Hafez-Torbati2018}%
  \BibitemOpen
  \bibfield  {author} {\bibinfo {author} {\bibfnamefont {M.}~\bibnamefont
  {Hafez-Torbati}}\ and\ \bibinfo {author} {\bibfnamefont {W.}~\bibnamefont
  {Hofstetter}},\ }\bibfield  {title} {\bibinfo {title} {Artificial {SU(3)}
  spin-orbit coupling and exotic {Mott} insulators},\ }\href
  {https://doi.org/10.1103/PhysRevB.98.245131} {\bibfield  {journal} {\bibinfo
  {journal} {Phys. Rev. B}\ }\textbf {\bibinfo {volume} {98}},\ \bibinfo
  {pages} {245131} (\bibinfo {year} {2018})}\BibitemShut {NoStop}%
\bibitem [{\citenamefont {Müller-Hartmann}(1989)}]{Mueller-Hartmann1989}%
  \BibitemOpen
  \bibfield  {author} {\bibinfo {author} {\bibfnamefont {E.}~\bibnamefont
  {Müller-Hartmann}},\ }\bibfield  {title} {\bibinfo {title} {Correlated
  fermions on a lattice in high dimensions},\ }\href
  {https://doi.org/10.1007/BF01311397} {\bibfield  {journal} {\bibinfo
  {journal} {Zeitschrift für Physik B Condensed Matter}\ }\textbf {\bibinfo
  {volume} {74}},\ \bibinfo {pages} {507} (\bibinfo {year} {1989})}\BibitemShut
  {NoStop}%
\bibitem [{\citenamefont {Vanhala}\ \emph {et~al.}(2016)\citenamefont
  {Vanhala}, \citenamefont {Siro}, \citenamefont {Liang}, \citenamefont
  {Troyer}, \citenamefont {Harju},\ and\ \citenamefont
  {Törmä}}]{Vanhala2016}%
  \BibitemOpen
  \bibfield  {author} {\bibinfo {author} {\bibfnamefont {T.~I.}\ \bibnamefont
  {Vanhala}}, \bibinfo {author} {\bibfnamefont {T.}~\bibnamefont {Siro}},
  \bibinfo {author} {\bibfnamefont {L.}~\bibnamefont {Liang}}, \bibinfo
  {author} {\bibfnamefont {M.}~\bibnamefont {Troyer}}, \bibinfo {author}
  {\bibfnamefont {A.}~\bibnamefont {Harju}},\ and\ \bibinfo {author}
  {\bibfnamefont {P.}~\bibnamefont {Törmä}},\ }\bibfield  {title} {\bibinfo
  {title} {Topological {Phase Transitions in the Repulsively Interacting
  {H}aldane-{H}ubbard Model}},\ }\href
  {https://doi.org/10.1103/PhysRevLett.116.225305} {\bibfield  {journal}
  {\bibinfo  {journal} {Phys. Rev. Lett.}\ }\textbf {\bibinfo {volume} {116}},\
  \bibinfo {pages} {225305} (\bibinfo {year} {2016})}\BibitemShut {NoStop}%
\bibitem [{\citenamefont {Mertz}\ \emph {et~al.}(2019)\citenamefont {Mertz},
  \citenamefont {Zantout},\ and\ \citenamefont {Valent\'{\i}}}]{Mertz2019}%
  \BibitemOpen
  \bibfield  {author} {\bibinfo {author} {\bibfnamefont {T.}~\bibnamefont
  {Mertz}}, \bibinfo {author} {\bibfnamefont {K.}~\bibnamefont {Zantout}},\
  and\ \bibinfo {author} {\bibfnamefont {R.}~\bibnamefont {Valent\'{\i}}},\
  }\bibfield  {title} {\bibinfo {title} {Statistical analysis of the {C}hern
  number in the interacting {H}aldane-{H}ubbard model},\ }\href
  {https://doi.org/10.1103/PhysRevB.100.125111} {\bibfield  {journal} {\bibinfo
   {journal} {Phys. Rev. B}\ }\textbf {\bibinfo {volume} {100}},\ \bibinfo
  {pages} {125111} (\bibinfo {year} {2019})}\BibitemShut {NoStop}%
\bibitem [{\citenamefont {Tupitsyn}\ and\ \citenamefont
  {Prokof'ev}(2019)}]{Tupitsyn2019}%
  \BibitemOpen
  \bibfield  {author} {\bibinfo {author} {\bibfnamefont {I.~S.}\ \bibnamefont
  {Tupitsyn}}\ and\ \bibinfo {author} {\bibfnamefont {N.~V.}\ \bibnamefont
  {Prokof'ev}},\ }\bibfield  {title} {\bibinfo {title} {Phase diagram topology
  of the {Haldane-Hubbard-Coulomb} model},\ }\href
  {https://doi.org/10.1103/PhysRevB.99.121113} {\bibfield  {journal} {\bibinfo
  {journal} {Phys. Rev. B}\ }\textbf {\bibinfo {volume} {99}},\ \bibinfo
  {pages} {121113} (\bibinfo {year} {2019})}\BibitemShut {NoStop}%
\bibitem [{\citenamefont {He}\ \emph {et~al.}(2024)\citenamefont {He},
  \citenamefont {Mondaini}, \citenamefont {Luo}, \citenamefont {Wang},\ and\
  \citenamefont {Hu}}]{He2024}%
  \BibitemOpen
  \bibfield  {author} {\bibinfo {author} {\bibfnamefont {W.-X.}\ \bibnamefont
  {He}}, \bibinfo {author} {\bibfnamefont {R.}~\bibnamefont {Mondaini}},
  \bibinfo {author} {\bibfnamefont {H.-G.}\ \bibnamefont {Luo}}, \bibinfo
  {author} {\bibfnamefont {X.}~\bibnamefont {Wang}},\ and\ \bibinfo {author}
  {\bibfnamefont {S.}~\bibnamefont {Hu}},\ }\bibfield  {title} {\bibinfo
  {title} {Phase transitions in the {H}aldane-{H}ubbard model},\ }\href
  {https://doi.org/10.1103/PhysRevB.109.035126} {\bibfield  {journal} {\bibinfo
   {journal} {Phys. Rev. B}\ }\textbf {\bibinfo {volume} {109}},\ \bibinfo
  {pages} {035126} (\bibinfo {year} {2024})}\BibitemShut {NoStop}%
\bibitem [{\citenamefont {Hafez-Torbati}(2025)}]{Hafez-Torbati2025}%
  \BibitemOpen
  \bibfield  {author} {\bibinfo {author} {\bibfnamefont {M.}~\bibnamefont
  {Hafez-Torbati}},\ }\bibfield  {title} {\bibinfo {title} {From explicit to
  spontaneous charge order and the fate of the antiferromagnetic quantum {H}all
  state},\ }\href {https://doi.org/10.1103/PhysRevB.111.125108} {\bibfield
  {journal} {\bibinfo  {journal} {Phys. Rev. B}\ }\textbf {\bibinfo {volume}
  {111}},\ \bibinfo {pages} {125108} (\bibinfo {year} {2025})}\BibitemShut
  {NoStop}%
\bibitem [{\citenamefont {Ishikawa}\ and\ \citenamefont
  {Matsuyama}(1987)}]{Ishikawa1987}%
  \BibitemOpen
  \bibfield  {author} {\bibinfo {author} {\bibfnamefont {K.}~\bibnamefont
  {Ishikawa}}\ and\ \bibinfo {author} {\bibfnamefont {T.}~\bibnamefont
  {Matsuyama}},\ }\bibfield  {title} {\bibinfo {title} {A microscopic theory of
  the quantum {H}all effect},\ }\href
  {https://doi.org/https://doi.org/10.1016/0550-3213(87)90160-X} {\bibfield
  {journal} {\bibinfo  {journal} {Nuclear Physics B}\ }\textbf {\bibinfo
  {volume} {280}},\ \bibinfo {pages} {523} (\bibinfo {year}
  {1987})}\BibitemShut {NoStop}%
\bibitem [{\citenamefont {Yoshida}\ \emph {et~al.}(2012)\citenamefont
  {Yoshida}, \citenamefont {Fujimoto},\ and\ \citenamefont
  {Kawakami}}]{Yoshida2012}%
  \BibitemOpen
  \bibfield  {author} {\bibinfo {author} {\bibfnamefont {T.}~\bibnamefont
  {Yoshida}}, \bibinfo {author} {\bibfnamefont {S.}~\bibnamefont {Fujimoto}},\
  and\ \bibinfo {author} {\bibfnamefont {N.}~\bibnamefont {Kawakami}},\
  }\bibfield  {title} {\bibinfo {title} {Correlation effects on a topological
  insulator at finite temperatures},\ }\href
  {https://doi.org/10.1103/PhysRevB.85.125113} {\bibfield  {journal} {\bibinfo
  {journal} {Phys. Rev. B}\ }\textbf {\bibinfo {volume} {85}},\ \bibinfo
  {pages} {125113} (\bibinfo {year} {2012})}\BibitemShut {NoStop}%
\bibitem [{\citenamefont {Irsigler}\ \emph {et~al.}(2021)\citenamefont
  {Irsigler}, \citenamefont {Grass}, \citenamefont {Zheng}, \citenamefont
  {Barbier},\ and\ \citenamefont {Hofstetter}}]{Irsigler2021}%
  \BibitemOpen
  \bibfield  {author} {\bibinfo {author} {\bibfnamefont {B.}~\bibnamefont
  {Irsigler}}, \bibinfo {author} {\bibfnamefont {T.}~\bibnamefont {Grass}},
  \bibinfo {author} {\bibfnamefont {J.-H.}\ \bibnamefont {Zheng}}, \bibinfo
  {author} {\bibfnamefont {M.}~\bibnamefont {Barbier}},\ and\ \bibinfo {author}
  {\bibfnamefont {W.}~\bibnamefont {Hofstetter}},\ }\bibfield  {title}
  {\bibinfo {title} {Topological {M}ott transition in a {W}eyl-{H}ubbard model:
  {D}ynamical mean-field theory study},\ }\href
  {https://doi.org/10.1103/PhysRevB.103.125132} {\bibfield  {journal} {\bibinfo
   {journal} {Phys. Rev. B}\ }\textbf {\bibinfo {volume} {103}},\ \bibinfo
  {pages} {125132} (\bibinfo {year} {2021})}\BibitemShut {NoStop}%
\bibitem [{\citenamefont {Manousakis}(1991)}]{Manousakis1991}%
  \BibitemOpen
  \bibfield  {author} {\bibinfo {author} {\bibfnamefont {E.}~\bibnamefont
  {Manousakis}},\ }\bibfield  {title} {\bibinfo {title} {The
  {Spin-$\frac{1}{2}$ Heisenberg Antiferromanget on a Square Lattice and its
  Application to the Cuprous Oxides}},\ }\href
  {https://journals.aps.org/rmp/abstract/10.1103/RevModPhys.63.1} {\bibfield
  {journal} {\bibinfo  {journal} {Rev. Mod. Phys.}\ }\textbf {\bibinfo {volume}
  {63}},\ \bibinfo {pages} {1} (\bibinfo {year} {1991})}\BibitemShut {NoStop}%
\bibitem [{\citenamefont {Rohringer}\ \emph {et~al.}(2018)\citenamefont
  {Rohringer}, \citenamefont {Hafermann}, \citenamefont {Toschi}, \citenamefont
  {Katanin}, \citenamefont {Antipov}, \citenamefont {Katsnelson}, \citenamefont
  {Lichtenstein}, \citenamefont {Rubtsov},\ and\ \citenamefont
  {Held}}]{Rohringer2018}%
  \BibitemOpen
  \bibfield  {author} {\bibinfo {author} {\bibfnamefont {G.}~\bibnamefont
  {Rohringer}}, \bibinfo {author} {\bibfnamefont {H.}~\bibnamefont
  {Hafermann}}, \bibinfo {author} {\bibfnamefont {A.}~\bibnamefont {Toschi}},
  \bibinfo {author} {\bibfnamefont {A.~A.}\ \bibnamefont {Katanin}}, \bibinfo
  {author} {\bibfnamefont {A.~E.}\ \bibnamefont {Antipov}}, \bibinfo {author}
  {\bibfnamefont {M.~I.}\ \bibnamefont {Katsnelson}}, \bibinfo {author}
  {\bibfnamefont {A.~I.}\ \bibnamefont {Lichtenstein}}, \bibinfo {author}
  {\bibfnamefont {A.~N.}\ \bibnamefont {Rubtsov}},\ and\ \bibinfo {author}
  {\bibfnamefont {K.}~\bibnamefont {Held}},\ }\bibfield  {title} {\bibinfo
  {title} {Diagrammatic routes to nonlocal correlations beyond dynamical mean
  field theory},\ }\href {https://doi.org/10.1103/RevModPhys.90.025003}
  {\bibfield  {journal} {\bibinfo  {journal} {Rev. Mod. Phys.}\ }\textbf
  {\bibinfo {volume} {90}},\ \bibinfo {pages} {025003} (\bibinfo {year}
  {2018})}\BibitemShut {NoStop}%
\bibitem [{\citenamefont {Van~{V}leck}(1932)}]{Vleck1932}%
  \BibitemOpen
  \bibfield  {author} {\bibinfo {author} {\bibfnamefont {J.~H.}\ \bibnamefont
  {Van~{V}leck}},\ }\href@noop {} {\emph {\bibinfo {title} {The {Theory of
  Electric and Magnetic Susceptibilities}}}}\ (\bibinfo  {publisher} {Oxford
  University Press},\ \bibinfo {year} {1932})\BibitemShut {NoStop}%
\bibitem [{\citenamefont {Garg}\ \emph {et~al.}(2006)\citenamefont {Garg},
  \citenamefont {Krishnamurthy},\ and\ \citenamefont {Randeria}}]{Garg2006}%
  \BibitemOpen
  \bibfield  {author} {\bibinfo {author} {\bibfnamefont {A.}~\bibnamefont
  {Garg}}, \bibinfo {author} {\bibfnamefont {H.~R.}\ \bibnamefont
  {Krishnamurthy}},\ and\ \bibinfo {author} {\bibfnamefont {M.}~\bibnamefont
  {Randeria}},\ }\bibfield  {title} {\bibinfo {title} {Can {Correlations Drive
  a Band Insulator Metallic?}},\ }\href
  {https://doi.org/10.1103/PhysRevLett.97.046403} {\bibfield  {journal}
  {\bibinfo  {journal} {Phys. Rev. Lett.}\ }\textbf {\bibinfo {volume} {97}},\
  \bibinfo {pages} {046403} (\bibinfo {year} {2006})}\BibitemShut {NoStop}%
\bibitem [{\citenamefont {Paris}\ \emph {et~al.}(2007)\citenamefont {Paris},
  \citenamefont {Bouadim}, \citenamefont {Hébert}, \citenamefont {Batrouni},\
  and\ \citenamefont {Scalettar}}]{Paris2007}%
  \BibitemOpen
  \bibfield  {author} {\bibinfo {author} {\bibfnamefont {N.}~\bibnamefont
  {Paris}}, \bibinfo {author} {\bibfnamefont {K.}~\bibnamefont {Bouadim}},
  \bibinfo {author} {\bibfnamefont {F.}~\bibnamefont {Hébert}}, \bibinfo
  {author} {\bibfnamefont {G.~G.}\ \bibnamefont {Batrouni}},\ and\ \bibinfo
  {author} {\bibfnamefont {R.~T.}\ \bibnamefont {Scalettar}},\ }\bibfield
  {title} {\bibinfo {title} {Quantum {M}onte~{Carlo Study of an
  Interaction-Driven Band-Insulator-to-Metal Transition}},\ }\href
  {https://doi.org/10.1103/PhysRevLett.98.046403} {\bibfield  {journal}
  {\bibinfo  {journal} {Phys. Rev. Lett.}\ }\textbf {\bibinfo {volume} {98}},\
  \bibinfo {pages} {046403} (\bibinfo {year} {2007})}\BibitemShut {NoStop}%
\bibitem [{\citenamefont {Byczuk}\ \emph {et~al.}(2009)\citenamefont {Byczuk},
  \citenamefont {Sekania}, \citenamefont {Hofstetter},\ and\ \citenamefont
  {Kampf}}]{Byczuk2009}%
  \BibitemOpen
  \bibfield  {author} {\bibinfo {author} {\bibfnamefont {K.}~\bibnamefont
  {Byczuk}}, \bibinfo {author} {\bibfnamefont {M.}~\bibnamefont {Sekania}},
  \bibinfo {author} {\bibfnamefont {W.}~\bibnamefont {Hofstetter}},\ and\
  \bibinfo {author} {\bibfnamefont {A.~P.}\ \bibnamefont {Kampf}},\ }\bibfield
  {title} {\bibinfo {title} {Insulating behavior with spin and charge order in
  the ionic {H}ubbard model},\ }\href
  {https://doi.org/10.1103/PhysRevB.79.121103} {\bibfield  {journal} {\bibinfo
  {journal} {Phys. Rev. B}\ }\textbf {\bibinfo {volume} {79}},\ \bibinfo
  {pages} {121103} (\bibinfo {year} {2009})}\BibitemShut {NoStop}%
\bibitem [{\citenamefont {Kancharla}\ and\ \citenamefont
  {Dagotto}(2007)}]{Kancharla2007}%
  \BibitemOpen
  \bibfield  {author} {\bibinfo {author} {\bibfnamefont {S.~S.}\ \bibnamefont
  {Kancharla}}\ and\ \bibinfo {author} {\bibfnamefont {E.}~\bibnamefont
  {Dagotto}},\ }\bibfield  {title} {\bibinfo {title} {Correlated {Insulated
  Phase Suggests Bond Order between Band and {M}ott Insulators in Two
  Dimensions}},\ }\href {https://doi.org/10.1103/PhysRevLett.98.016402}
  {\bibfield  {journal} {\bibinfo  {journal} {Phys. Rev. Lett.}\ }\textbf
  {\bibinfo {volume} {98}},\ \bibinfo {pages} {016402} (\bibinfo {year}
  {2007})}\BibitemShut {NoStop}%
\bibitem [{\citenamefont {Hafez-Torbati}\ and\ \citenamefont
  {Uhrig}(2016)}]{Hafez-Torbati2016}%
  \BibitemOpen
  \bibfield  {author} {\bibinfo {author} {\bibfnamefont {M.}~\bibnamefont
  {Hafez-Torbati}}\ and\ \bibinfo {author} {\bibfnamefont {G.~S.}\ \bibnamefont
  {Uhrig}},\ }\bibfield  {title} {\bibinfo {title} {Orientational bond and
  {N}\'eel order in the two-dimensional ionic {H}ubbard model},\ }\href
  {https://doi.org/10.1103/PhysRevB.93.195128} {\bibfield  {journal} {\bibinfo
  {journal} {Phys. Rev. B}\ }\textbf {\bibinfo {volume} {93}},\ \bibinfo
  {pages} {195128} (\bibinfo {year} {2016})}\BibitemShut {NoStop}%
\bibitem [{\citenamefont {Khomskii}\ and\ \citenamefont
  {Streltsov}(2021)}]{Khomskii2021}%
  \BibitemOpen
  \bibfield  {author} {\bibinfo {author} {\bibfnamefont {D.~I.}\ \bibnamefont
  {Khomskii}}\ and\ \bibinfo {author} {\bibfnamefont {S.~V.}\ \bibnamefont
  {Streltsov}},\ }\bibfield  {title} {\bibinfo {title} {Orbital {Effects in
  Solids: Basics, Recent Progress, and Opportunities}},\ }\href
  {https://doi.org/10.1021/acs.chemrev.0c00579} {\bibfield  {journal} {\bibinfo
   {journal} {Chem. Rev.}\ }\textbf {\bibinfo {volume} {121}},\ \bibinfo
  {pages} {2992} (\bibinfo {year} {2021})}\BibitemShut {NoStop}%
\bibitem [{\citenamefont {Liu}\ \emph {et~al.}(2011)\citenamefont {Liu},
  \citenamefont {Jiang},\ and\ \citenamefont {Yao}}]{Liu2011}%
  \BibitemOpen
  \bibfield  {author} {\bibinfo {author} {\bibfnamefont {C.-C.}\ \bibnamefont
  {Liu}}, \bibinfo {author} {\bibfnamefont {H.}~\bibnamefont {Jiang}},\ and\
  \bibinfo {author} {\bibfnamefont {Y.}~\bibnamefont {Yao}},\ }\bibfield
  {title} {\bibinfo {title} {Low-energy effective {Hamiltonian} involving
  spin-orbit coupling in silicene and two-dimensional germanium and tin},\
  }\href {https://doi.org/10.1103/PhysRevB.84.195430} {\bibfield  {journal}
  {\bibinfo  {journal} {Phys. Rev. B}\ }\textbf {\bibinfo {volume} {84}},\
  \bibinfo {pages} {195430} (\bibinfo {year} {2011})}\BibitemShut {NoStop}%
\bibitem [{\citenamefont {Ni}\ \emph {et~al.}(2012)\citenamefont {Ni},
  \citenamefont {Liu}, \citenamefont {Tang}, \citenamefont {Zheng},
  \citenamefont {Zhou}, \citenamefont {Qin}, \citenamefont {Gao}, \citenamefont
  {Yu},\ and\ \citenamefont {Lu}}]{Ni2012}%
  \BibitemOpen
  \bibfield  {author} {\bibinfo {author} {\bibfnamefont {Z.}~\bibnamefont
  {Ni}}, \bibinfo {author} {\bibfnamefont {Q.}~\bibnamefont {Liu}}, \bibinfo
  {author} {\bibfnamefont {K.}~\bibnamefont {Tang}}, \bibinfo {author}
  {\bibfnamefont {J.}~\bibnamefont {Zheng}}, \bibinfo {author} {\bibfnamefont
  {J.}~\bibnamefont {Zhou}}, \bibinfo {author} {\bibfnamefont {R.}~\bibnamefont
  {Qin}}, \bibinfo {author} {\bibfnamefont {Z.}~\bibnamefont {Gao}}, \bibinfo
  {author} {\bibfnamefont {D.}~\bibnamefont {Yu}},\ and\ \bibinfo {author}
  {\bibfnamefont {J.}~\bibnamefont {Lu}},\ }\bibfield  {title} {\bibinfo
  {title} {Tunable bandgap in silicene and germanene},\ }\href
  {https://doi.org/10.1021/nl203065e} {\bibfield  {journal} {\bibinfo
  {journal} {Nano Lett.}\ }\textbf {\bibinfo {volume} {12}},\ \bibinfo {pages}
  {113} (\bibinfo {year} {2012})}\BibitemShut {NoStop}%
\bibitem [{\citenamefont {Zhang}\ \emph {et~al.}(2009)\citenamefont {Zhang},
  \citenamefont {Tang}, \citenamefont {Girit}, \citenamefont {Hao},
  \citenamefont {Martin}, \citenamefont {Zettl}, \citenamefont {Crommie},
  \citenamefont {Shen},\ and\ \citenamefont {Wang}}]{Zhang2009a}%
  \BibitemOpen
  \bibfield  {author} {\bibinfo {author} {\bibfnamefont {Y.}~\bibnamefont
  {Zhang}}, \bibinfo {author} {\bibfnamefont {T.-T.}\ \bibnamefont {Tang}},
  \bibinfo {author} {\bibfnamefont {C.}~\bibnamefont {Girit}}, \bibinfo
  {author} {\bibfnamefont {Z.}~\bibnamefont {Hao}}, \bibinfo {author}
  {\bibfnamefont {M.~C.}\ \bibnamefont {Martin}}, \bibinfo {author}
  {\bibfnamefont {A.}~\bibnamefont {Zettl}}, \bibinfo {author} {\bibfnamefont
  {M.~F.}\ \bibnamefont {Crommie}}, \bibinfo {author} {\bibfnamefont {Y.~R.}\
  \bibnamefont {Shen}},\ and\ \bibinfo {author} {\bibfnamefont
  {F.}~\bibnamefont {Wang}},\ }\bibfield  {title} {\bibinfo {title} {Direct
  observation of a widely tunable bandgap in bilayer graphene},\ }\href
  {https://doi.org/10.1038/nature08105} {\bibfield  {journal} {\bibinfo
  {journal} {Nature}\ }\textbf {\bibinfo {volume} {459}},\ \bibinfo {pages}
  {820} (\bibinfo {year} {2009})}\BibitemShut {NoStop}%
\bibitem [{\citenamefont {Sakanashi}\ \emph {et~al.}(2021)\citenamefont
  {Sakanashi}, \citenamefont {Wada}, \citenamefont {Murase}, \citenamefont
  {Oto}, \citenamefont {Kim}, \citenamefont {Watanabe}, \citenamefont
  {Taniguchi}, \citenamefont {Bird}, \citenamefont {Ferry},\ and\ \citenamefont
  {Aoki}}]{Sakanashi2021}%
  \BibitemOpen
  \bibfield  {author} {\bibinfo {author} {\bibfnamefont {K.}~\bibnamefont
  {Sakanashi}}, \bibinfo {author} {\bibfnamefont {N.}~\bibnamefont {Wada}},
  \bibinfo {author} {\bibfnamefont {K.}~\bibnamefont {Murase}}, \bibinfo
  {author} {\bibfnamefont {K.}~\bibnamefont {Oto}}, \bibinfo {author}
  {\bibfnamefont {G.-H.}\ \bibnamefont {Kim}}, \bibinfo {author} {\bibfnamefont
  {K.}~\bibnamefont {Watanabe}}, \bibinfo {author} {\bibfnamefont
  {T.}~\bibnamefont {Taniguchi}}, \bibinfo {author} {\bibfnamefont {J.~P.}\
  \bibnamefont {Bird}}, \bibinfo {author} {\bibfnamefont {D.~K.}\ \bibnamefont
  {Ferry}},\ and\ \bibinfo {author} {\bibfnamefont {N.}~\bibnamefont {Aoki}},\
  }\bibfield  {title} {\bibinfo {title} {Valley polarized conductance
  quantization in bilayer graphene narrow quantum point contact},\ }\href
  {https://doi.org/10.1063/5.0052845} {\bibfield  {journal} {\bibinfo
  {journal} {Applied Physics Letters}\ }\textbf {\bibinfo {volume} {118}},\
  \bibinfo {pages} {263102} (\bibinfo {year} {2021})}\BibitemShut {NoStop}%
\bibitem [{\citenamefont {Asai}\ \emph {et~al.}(2016)\citenamefont {Asai},
  \citenamefont {Soda}, \citenamefont {Kasatani}, \citenamefont {Ono},
  \citenamefont {Avdeev},\ and\ \citenamefont {Masuda}}]{Asai2016}%
  \BibitemOpen
  \bibfield  {author} {\bibinfo {author} {\bibfnamefont {S.}~\bibnamefont
  {Asai}}, \bibinfo {author} {\bibfnamefont {M.}~\bibnamefont {Soda}}, \bibinfo
  {author} {\bibfnamefont {K.}~\bibnamefont {Kasatani}}, \bibinfo {author}
  {\bibfnamefont {T.}~\bibnamefont {Ono}}, \bibinfo {author} {\bibfnamefont
  {M.}~\bibnamefont {Avdeev}},\ and\ \bibinfo {author} {\bibfnamefont
  {T.}~\bibnamefont {Masuda}},\ }\bibfield  {title} {\bibinfo {title} {Magnetic
  ordering of the buckled honeycomb lattice antiferromagnet
  {Ba$_2$NiTeO$_6$}},\ }\href {https://doi.org/10.1103/PhysRevB.93.024412}
  {\bibfield  {journal} {\bibinfo  {journal} {Phys. Rev. B}\ }\textbf {\bibinfo
  {volume} {93}},\ \bibinfo {pages} {024412} (\bibinfo {year}
  {2016})}\BibitemShut {NoStop}%
\bibitem [{\citenamefont {Asai}\ \emph {et~al.}(2017)\citenamefont {Asai},
  \citenamefont {Soda}, \citenamefont {Kasatani}, \citenamefont {Ono},
  \citenamefont {Garlea}, \citenamefont {Winn},\ and\ \citenamefont
  {Masuda}}]{Asai2017}%
  \BibitemOpen
  \bibfield  {author} {\bibinfo {author} {\bibfnamefont {S.}~\bibnamefont
  {Asai}}, \bibinfo {author} {\bibfnamefont {M.}~\bibnamefont {Soda}}, \bibinfo
  {author} {\bibfnamefont {K.}~\bibnamefont {Kasatani}}, \bibinfo {author}
  {\bibfnamefont {T.}~\bibnamefont {Ono}}, \bibinfo {author} {\bibfnamefont
  {V.~O.}\ \bibnamefont {Garlea}}, \bibinfo {author} {\bibfnamefont
  {B.}~\bibnamefont {Winn}},\ and\ \bibinfo {author} {\bibfnamefont
  {T.}~\bibnamefont {Masuda}},\ }\bibfield  {title} {\bibinfo {title} {Spin
  dynamics in the stripe-ordered buckled honeycomb lattice antiferromagnet
  {Ba$_2$NiTeO$_6$}},\ }\href {https://doi.org/10.1103/PhysRevB.96.104414}
  {\bibfield  {journal} {\bibinfo  {journal} {Phys. Rev. B}\ }\textbf {\bibinfo
  {volume} {96}},\ \bibinfo {pages} {104414} (\bibinfo {year}
  {2017})}\BibitemShut {NoStop}%
\bibitem [{\citenamefont {Thakur}\ \emph {et~al.}(2022)\citenamefont {Thakur},
  \citenamefont {Hansen}, \citenamefont {Schnelle}, \citenamefont {Guo},
  \citenamefont {Janson}, \citenamefont {van~den Brink}, \citenamefont
  {Felser},\ and\ \citenamefont {Jansen}}]{Thakur2022}%
  \BibitemOpen
  \bibfield  {author} {\bibinfo {author} {\bibfnamefont {G.~S.}\ \bibnamefont
  {Thakur}}, \bibinfo {author} {\bibfnamefont {T.~C.}\ \bibnamefont {Hansen}},
  \bibinfo {author} {\bibfnamefont {W.}~\bibnamefont {Schnelle}}, \bibinfo
  {author} {\bibfnamefont {S.}~\bibnamefont {Guo}}, \bibinfo {author}
  {\bibfnamefont {O.}~\bibnamefont {Janson}}, \bibinfo {author} {\bibfnamefont
  {J.}~\bibnamefont {van~den Brink}}, \bibinfo {author} {\bibfnamefont
  {C.}~\bibnamefont {Felser}},\ and\ \bibinfo {author} {\bibfnamefont
  {M.}~\bibnamefont {Jansen}},\ }\bibfield  {title} {\bibinfo {title} {Buckled
  honeycomb lattice compound {Sr$_3$CaOs$_2$O$_9$} exhibiting
  antiferromagnetism above room temperature},\ }\href
  {https://doi.org/10.1021/acs.chemmater.2c00740} {\bibfield  {journal}
  {\bibinfo  {journal} {Chem. Mater.}\ }\textbf {\bibinfo {volume} {34}},\
  \bibinfo {pages} {4741} (\bibinfo {year} {2022})}\BibitemShut {NoStop}%
\bibitem [{\citenamefont {Xiao}\ \emph {et~al.}(2011)\citenamefont {Xiao},
  \citenamefont {Zhu}, \citenamefont {Ran}, \citenamefont {Nagaosa},\ and\
  \citenamefont {Okamoto}}]{Xiao2011}%
  \BibitemOpen
  \bibfield  {author} {\bibinfo {author} {\bibfnamefont {D.}~\bibnamefont
  {Xiao}}, \bibinfo {author} {\bibfnamefont {W.}~\bibnamefont {Zhu}}, \bibinfo
  {author} {\bibfnamefont {Y.}~\bibnamefont {Ran}}, \bibinfo {author}
  {\bibfnamefont {N.}~\bibnamefont {Nagaosa}},\ and\ \bibinfo {author}
  {\bibfnamefont {S.}~\bibnamefont {Okamoto}},\ }\bibfield  {title} {\bibinfo
  {title} {Interface engineering of quantum {H}all effects in digital
  transition metal oxide heterostructures},\ }\href
  {https://doi.org/10.1038/ncomms1602} {\bibfield  {journal} {\bibinfo
  {journal} {Nature Communications}\ }\textbf {\bibinfo {volume} {2}},\
  \bibinfo {pages} {596} (\bibinfo {year} {2011})}\BibitemShut {NoStop}%
\bibitem [{\citenamefont {Liang}\ \emph {et~al.}(2013)\citenamefont {Liang},
  \citenamefont {Wu},\ and\ \citenamefont {Hu}}]{Liang2013}%
  \BibitemOpen
  \bibfield  {author} {\bibinfo {author} {\bibfnamefont {Q.-F.}\ \bibnamefont
  {Liang}}, \bibinfo {author} {\bibfnamefont {L.-H.}\ \bibnamefont {Wu}},\ and\
  \bibinfo {author} {\bibfnamefont {X.}~\bibnamefont {Hu}},\ }\bibfield
  {title} {\bibinfo {title} {Electrically tunable topological state in [111]
  perovskite materials with an antiferromagnetic exchange field},\ }\href
  {https://doi.org/10.1088/1367-2630/15/6/063031} {\bibfield  {journal}
  {\bibinfo  {journal} {New Journal of Physics}\ }\textbf {\bibinfo {volume}
  {15}},\ \bibinfo {pages} {063031} (\bibinfo {year} {2013})}\BibitemShut
  {NoStop}%
\end{thebibliography}

%

\end{document}


\newcommand{\be}{\begin{equation}}
\newcommand{\ee}{\end{equation}}
\newcommand{\bearr}{\begin{eqnarray}}
\newcommand{\eearr}{\end{eqnarray}}
\newcommand{\bseq}{\begin{subequations}}
\newcommand{\eseq}{\end{subequations}}
\newcommand{\nn}{\nonumber}
\newcommand{\dagg}{{\dagger}}
\newcommand{\vpdag}{{\vphantom{\dagger}}}
\newcommand{\vecr}{\vec{r}}
\newcommand{\bs}{\boldsymbol}
\newcommand{\up}{\uparrow}
\newcommand{\down}{\downarrow}
\newcommand{\fns}{\footnotesize}
\newcommand{\ns}{\normalsize}
\newcommand{\cdag}{c^{\dagger}}
\newcommand{\so}{\lambda_{\rm SO}}
\newcommand{\jh}{J_{\rm H}}
\newcommand{\tn}{T_{\rm N}}
\newcommand{\tq}{T_{q}}
\newcommand{\la}{\langle}
\newcommand{\ra}{\rangle}
\newcommand{\sgn}{\text{sgn}}

\title{Supplemental Material: High-Temperature Quantum Anomalous Hall Effect\\in Buckled Honeycomb Antiferromagnets}

\author{Mohsen Hafez-Torbati}
\email{m.hafeztorbati@gmail.com}
\affiliation{Department of Physics, Shahid Beheshti University, 1983969411, Tehran, Iran}

\author{G\"otz S. Uhrig}
\email{goetz.uhrig@tu-dortmund.de}
\affiliation{Condensed Matter Theory, Department of Physics, TU Dortmund University, 44221 Dortmund, Germany}

\maketitle

\section{dynamical mean-field theory}
\label{sec:dmft}

As mentioned in the main text we employ the real-space realization of the dynamical mean-field theory (DMFT)
\cite{Potthoff1999,Song2008,Snoek2008}, specifically
the implementation introduced in Ref. \cite{Hafez-Torbati2018}. The application to a generalized Kondo
lattice model of the form of Eq. (2) in the main text is already discussed in details in Ref. \cite{Hafez-Torbati2022}
and also in Ref. \cite{Hafez-Torbati2021}. In the study of the bulk properties,
i.e., the periodic boundary conditions in both directions,
the impurity model is set up for two representative lattice sites corresponding to the two sublattices.

In the DMFT approximation, the non-local interactions are substituted with their Hartree counterparts \cite{Mueller-Hartmann1989}.
Assuming a uniform N\'eel AF order, a mean-field decoupling simplifies the non-local Heisenberg interactions in Eq. (2)
to the effective magnetic fields
\bseq
\label{eq:h}
\begin{align}
h_i^{\rm loc}&=Z_i \langle S_i^z +s_i^z \rangle \ , \\
h_i^{\rm iti}&=Z_i \langle S_i^z \rangle \ ,
\end{align}
\eseq
acting respectively on the localized $S_i^z$ and the itinerant $s_i^z$ spins at the lattice site $i$
\cite{Hafez-Torbati2022,Hafez-Torbati2021}.
$Z_i$ denotes the coordination number for the lattice site $i$. For the periodic boundary conditions in both directions
one simply has $Z_i=3$ independent of $i$. For the cylindrical geometry, see Fig. 4 in the main text, $Z_i=2$ if $i$ is located
at the edges and $Z_i=3$ otherwise. We recall that the Heisenberg interaction in Eq. (2) in the main text is counted only once
on each lattice bond.

The uniform N\'eel AF order assumed in the derivation of Eq. \eqref{eq:h} is justified for the cylindrical
geometry because the edge effects on $\la S_i^z \ra$ and $\la s_i^z \ra$ is extremely small.
To study the system with the cylindrical geometry we fix the expectation values
$\la S_i^z \ra$  and $\la s_i^z \ra$ in Eq. \eqref{eq:h} to what we have already found for the bulk, using
the periodic boundary conditions in both directions. This fixes the Hamiltonian and avoids
too many unknown parameters, which allows
for an easier and faster convergence of the DMFT loop \cite{Hafez-Torbati2025}.
Note that for the cylindrical geometry, the impurity model at each DMFT iteration
has to be set up and solved for the number of lattice sites $2N_x=160$ in the unit cell, see the dashed box
in Fig. 4 in the main text. The edge effects are taken into account via the hopping term, the spin-orbit coupling term,
the self-energy,
and the coordination number $Z_i$ in Eq. \eqref{eq:h}. The self-energy is computed at the $2N_x=160$ different
lattice sites in the unit cell.

\begin{figure}[t]
   \begin{center}
   \includegraphics[width=0.26\textwidth,angle=-90]{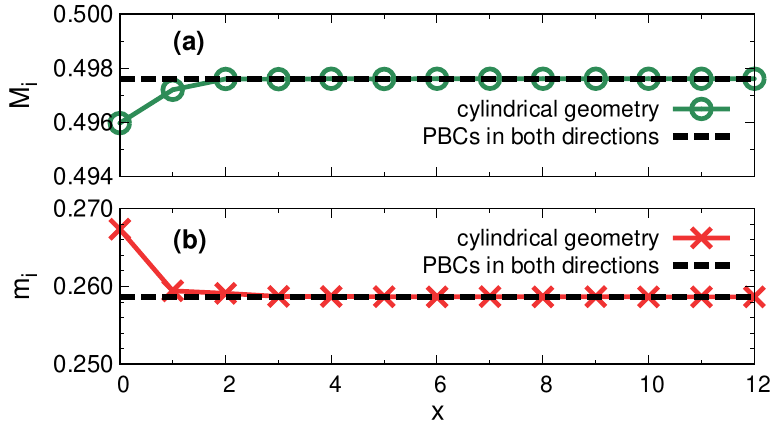}
   \caption{Position dependence of the local magnetizations $M_i=|\la S_i^z \ra|$ (a) and $m_i=|\la s_i^z \ra|$ (b) for the
cylindrical geometry sketched in Fig. 4 in the main text. The horizontal lines denote the values obtained using
the periodic boundary conditions (PBCs) in both directions. The results are for the same model parameters
as in Fig. 5 in the main text at $T=0.1t$.}
   \label{fig:spin}
   \end{center}
\end{figure}

Figure \ref{fig:spin} represents the position dependence of the local magnetizations $M_i=|\la S_i^z \ra|$ and
$m_i=|\la s_i^z \ra|$ for the cylindrical geometry sketched in Fig. 4 in the main text.
The results are for the same model parameters as in Fig. 5 in the main text at $T=0.1t$.
The horizontal dashed lines
in Fig. \ref{fig:spin} denote the values obtained using the periodic boundary conditions in both directions.
The extremely small edge effects on $\la S_i^z \ra$ and $\la s_i^z \ra$ justify our assumption of the
uniform N\'eel AF order in the derivation of Eq. \eqref{eq:h} for the cylindrical geometry.

\section{ionicity}
The Kondo lattice approximation, in Eq. (2) in the main text, of the multi-orbital Hubbard model
limits the possible amount of charge transfer between the higher and the lower energy sublattices.
To examine the amount of charge transfer between the two sublattices we consider the ionicity
\be
I=\frac{\langle n_{B}\rangle -\langle n_{A} \rangle}{2}
\label{eq:ion}
\ee
where $\langle n_{A}\rangle$ and $\langle n_{B}\rangle$ represent the electron density on
the higher energy $A$ and the lower energy $B$ sublattices, respectively.

\begin{figure}[t]
   \begin{center}
   \includegraphics[width=0.21\textwidth,angle=-90]{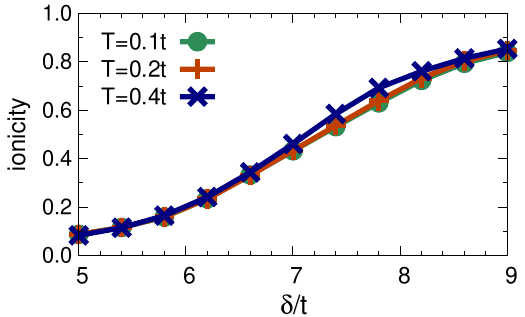}
   \caption{Ionicity vs the alternating sublattice potential $\delta$ at different temperatures.
   The results are for the same model parameters as in Fig. 2 in the main text.}
   \label{fig:ion}
   \end{center}
\end{figure}

We plotted the ionicity vs $\delta$ at different temperatures in Fig. \ref{fig:ion}.
The results are for the same model parameters as in Fig. 2
in the main text. For the intermediate values of the alternating sublattice potential $\delta \sim 7t$,
which we consider particularly, the ionicity takes a value of about $0.4$. Such a value, still away from the saturation
limit of $1$, justifies the Kondo lattice approximation of the multi-orbital Hubbard model. For $\delta\gtrsim 8t$, one can
see a precursor of saturation in the ionicity and hence more than a single itinerant orbital is expected
to be required for an accurate description of the system.

\section{Hall Conductance}
The topological invariant form of the Hall conductance $\sigma_{xy}$ in Eq. (3) in the main text
can be simplified to
\bseq
\label{eq:hall}
\begin{align}
\label{eq:hall:a}
\sigma_{xy}&= \sum_{\alpha} \sigma_{xy}^\alpha= \sum_{\alpha}\sum_n \mathcal{S}^\alpha_{xy}({\rm i}\omega_n) \ , \\
\mathcal{S}^\alpha_{xy}({\rm i}\omega_n)&= \frac{e^2}{h}\frac{T}{2\pi}
{\rm Re}\!\left[   \int \!{\rm d}\vec{k}
~{\rm Tr}\!\left[
\boldsymbol{G}^\vpdag_{\!\alpha}~ \partial_{k_y}\! \boldsymbol{\mathcal{H}}^{(0)}_\alpha
\right.
\right. \nn \\
\label{eq:hall:b}
&\quad\quad\quad\quad\quad\quad\quad\quad \times
\left.
\left.
\!\!\partial_{\omega_{n}}\! \boldsymbol{G}^\vpdag_{\!\alpha}
~\partial_{k_x}\! \boldsymbol{\mathcal{H}}^{(0)}_\alpha
\right]
\vphantom{\sum_\alpha\int}
\right]
\end{align}
\eseq
where we have used the facts that the conductance is an antisymmetric tensor and the self-energy in the DMFT approximation
is momentum-independent \cite{Yoshida2012}.
The sum is over the Matsubara frequencies $\omega_n=(2n+1)\pi T$, the momentum integration is over the first Brillouin zone
of the honeycomb lattice, $\boldsymbol{\mathcal{H}}^{(0)}_\alpha \equiv \boldsymbol{\mathcal{H}}^{(0)}_\alpha(\vec{k})$ is
the $2\times2$ Bloch Hamiltonian matrix, and
$\partial_{\omega_n} \boldsymbol{G}_\alpha \equiv \partial_{\omega} \boldsymbol{G}_\alpha({\rm i}\omega,\vec{k})|_{\omega=\omega_n}$.
The Bloch Hamiltonian matrix and its derivatives in Eq. \eqref{eq:hall:b} can be calculated analytically.
The Green's function can also be rigorously evaluated at different frequencies using the self-energy obtained
from the DMFT. What is somewhat challenging is an accurate estimate of the derivative of the Green's function
$\partial_{\omega_n} \boldsymbol{G}_\alpha$ \cite{Yoshida2012,Irsigler2021} which requires analytic continuation.

\begin{figure}[t]
   \begin{center}
   \includegraphics[width=0.47\textwidth,angle=0]{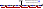}
   \caption{Schematic comparison of Matsubara and fictitious frequencies for $T_{\rm fic}=T/5$. Under brackets
   specify the frequencies used in Eq. \eqref{eq:diff} to compute the derivative of the Green's function
   at a Matsubara frequency.}
   \label{fig:diff}
   \end{center}
\end{figure}

We perform the DMFT loop for the fictitious
frequencies $\omega_n^{\rm fic}=(2n+1)\pi T_{\rm fic}$
with $T_{\rm fic}=T/(2l+1)$. The odd number $2l+1$ is to guarantee that the fictitious frequencies
involve the Matsubara frequencies. In addition, there are $2l$ additional fictitious frequencies between
each two successive Matsubara frequencies, see Fig. \ref{fig:diff}.
To accurately estimate the derivative of the Green's function at the Matsubara frequencies
we use the five-point central difference formula,
\begin{align}
\label{eq:diff}
\partial_\omega \boldsymbol{G}({\rm i}\omega) \big{|}_{\omega=\omega_n^{\rm fic}}
=&\frac{1}{{24\pi T_{\rm fic}}} \left(
\boldsymbol{G}({\rm i}\omega_{n-2}^{\rm fic})+8\boldsymbol{G}({\rm i}\omega_{n+1}^{\rm fic})
\right. \nn \\
&\quad\quad\left.
-8\boldsymbol{G}({\rm i}\omega_{n-1}^{\rm fic})-\boldsymbol{G}({\rm i}\omega_{n+2}^{\rm fic}) \right)
\end{align}
where we have dropped the spin $\alpha$ and the momentum $\vec{k}$ dependence of the Green's function
to lighten the notation. The fictitious frequency $\omega_n^{\rm fic}$, at which the derivative is calculated,
is assumed to match a Matsubara frequency.
The frequencies used in Eq. \eqref{eq:diff} to compute the derivative of the Green's function
at a Matsubara frequency are distinguished by an under bracket in Fig. \ref{fig:diff} for $T_{\rm fic}=T/5$.
We have mainly used $T_{\rm fic}=T/5$ in the calculations. Nevertheless, for selective points close to the phase transitions
we have checked that the results accurately match the results obtained for $T_{\rm fic}=T/3$ and
$T/7$. For example, for the model parameters in Fig. 2 in the main text at $T=0.02$ and $\delta=6.3t$
(just above the transition point $\delta_{c_1} \simeq 6.2t$, see also Fig. \ref{fig:T0.02}(b) discussed in the next section)
with $n_b=6$ we
find $h\sigma_{yx}/e^2 \simeq 0.9897$, $0.9993$, $0.9991$, and
$0.9992$, respectively, for $T_{\rm fic}=T$, $T/3$, $T/5$, and $T/7$.

\begin{figure}[t]
   \begin{center}
   \includegraphics[width=0.3\textwidth,angle=-90]{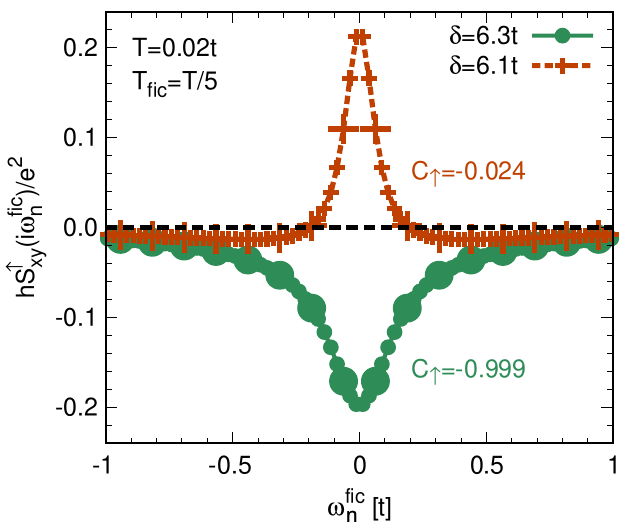}
   \caption{The summand $\mathcal{S}^\up_{xy}({\rm i}\omega^{\rm fic}_n)$ defined by Eq. \eqref{eq:hall} vs
   the fictitious frequency $\omega^{\rm fic}_n$ for $\delta=6.1t$ and $6.3t$. The larger symbols distinguish the
   values at Matsubara frequencies. The results are for the same model parameters
   as in Fig. 2 in the main text at $T=0.02t$. The fictitious temperature $T_{\rm fic}=T/5$ and the number of bath sites
   $n_b=6$ are used in the calculations. The Chern number $\mathcal{C}_{\alpha}:=h\sigma_{xy}^\alpha/e^2$
   is obtained by summing up the contributions at Matsubara frequencies.}
   \label{fig:summand}
   \end{center}
\end{figure}

It is interesting to see how the summand in Eq. \eqref{eq:hall} changes across a topological phase transition,
where the Hall conductance abruptly jumps between $0$ and $e^2/h$.
We consider the topological phase transition at $\delta_{c_1} \simeq 6.2t$ at the low temperature $T=0.02t$
in Fig. 2 in the main text, see also Fig. \ref{fig:T0.02}(b) discussed in the next section.
In Fig. \ref{fig:summand} we plotted $\mathcal{S}^\up_{xy}({\rm i}\omega^{\rm fic}_n)$ vs $\omega^{\rm fic}_n$
for $\delta=6.1t$ and $6.3t$.
For the solution with the magnetization on the higher-energy sublattice pointing in the positive $z$ direction
the topological phase transition is due to the spin $\up$.
The spin $\down$ always remains topologically trivial.
We used $T_{\rm fic}=T/5$ and the number of bath sites $n_b=6$ in the calculations.
The larger symbols in Fig. \ref{fig:summand} correspond to the values of $\mathcal{S}^\up_{xy}({\rm i}\omega^{\rm fic}_n)$
at Matsubara frequencies. Note that the Chern number $\mathcal{C}_{\alpha}:=h\sigma_{xy}^\alpha/e^2$ is obtained by
summing up the values of $\mathcal{S}^{\alpha}_{xy}({\rm i}\omega^{\rm fic}_n)$ at
{\it Matsubara} frequencies according to Eq. \eqref{eq:hall}.
Fig. \ref{fig:summand} illustrates how the different contributions in the trivial region cancel out, while in the
topological region they add up to a finite quantized value.

The topological Hamiltonian method \cite{Wang2012} provides us with a simple way to check our results for the Hall conductance
obtained via Eq. \eqref{eq:hall} at low temperatures.
The topological properties of an interacting system at zero temperature can be determined via an effective
non-interacting model known as the topological Hamiltonian.
Definitely, the method has some limitations. For example, it holds for fermionic systems at $T=0$ and
cannot be applied to bosonic bands \cite{Hawashin2024}.
The topological Hamiltonian
is given by the non-interacting part of the original Hamiltonian plus the self-energy at
the zero frequency \cite{Wang2012}. Since the self-energy in the DMFT is local, its effect will be only to modify the onsite energies.
Similar to Ref. \cite{Hafez-Torbati2024}, the topological Hamiltonian for our system is given by the
Kane-Mele model (corresponding to Eq. (2) in the main text with no localized spin and no Hubbard $U$)
with the effective spin-dependent alternating sublattice potential
\be
\tilde{\delta}_\alpha=\delta-\frac{3}{2}{\rm sgn}(\alpha) \la S_{A}^z \ra J
+ \frac{\Sigma_{\alpha,A}(0)-\Sigma_{\alpha,B}(0)}{2} \quad,
\ee
where we have defined ${\rm sgn}(\up)=+1$ and ${\rm sgn}(\down)=-1$, and $A$ and $B$ represent
the higher- and the lower-energy sublattices, respectively.
The spin component $\alpha$ is in the quantum Hall state with $\sigma_{yx}^\alpha={\rm sgn}(\alpha)e^2/h$
for $|\tilde{\delta}_\alpha|<3\sqrt{3}\so$ and in the trivial state with
$\sigma_{yx}^\alpha=0$ for $|\tilde{\delta}_\alpha|>3\sqrt{3}\so$. The topological phase transitions that we find
based on the calculation of the Hall conductance in Eq. \eqref{eq:hall} at low temperatures
is always checked to be in perfect agreement with the results of the topological Hamiltonian approach, see
the next section for an example.

\section{Further details of the phase diagram}

\begin{figure}[t]
   \begin{center}
   \includegraphics[width=0.42\textwidth,angle=-90]{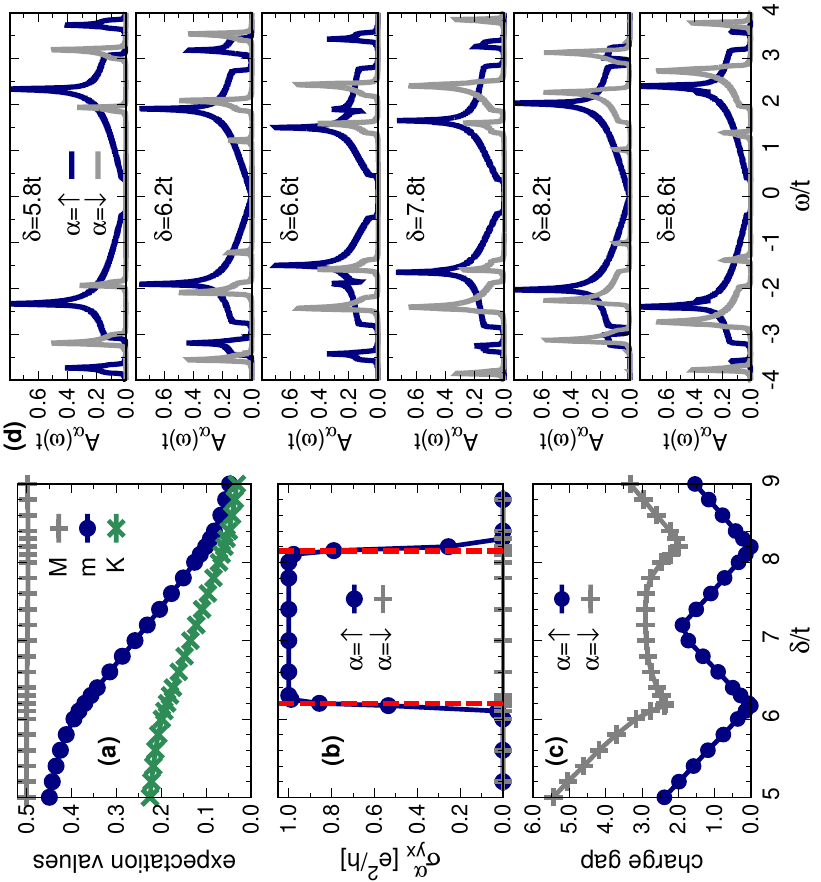}
   \caption{Local magnetization of the itinerant electrons $m=|\la s_i^z \ra|$ and the localized spins $M=|\la S_i^z \ra|$
   and the local Kondo correlation $K=\la \vec{S}_i \cdot \vec{s}_i\ra$ (a), individual spin contributions to the Hall
   conductance $\sigma_{yx}^\alpha$ (b), and the charge gap for spin $\up$ and $\down$ (c) plotted vs the alternating
   sublattice potential $\delta$.
   (d) Local spectral function vs frequency for
   different values of $\delta$ across the topological phase transitions $\delta_{c_1} \simeq 6.2t$ and $\delta_{c_2} \simeq 8.2t$.
   The results are for $T=0.02t$ and the model parameters $S=1/2$, $U=12t$, $\jh=0.2U$,
   $J=4t^2/\Delta_0=0.2\bar{7}t$, and $\so=0.2t$. The number of bath sites $n_b=6$ is used in the ED impurity solver.
   The red vertical dashed lines in
   (b) denote the location of the topological phase transitions predicted by the topological Hamiltonian approach for $T=0$.}
   \label{fig:T0.02}
   \end{center}
\end{figure}

In this section we provide further details leading to the phase diagram in Fig. 2 in the main text.
Figure \ref{fig:T0.02} depicts results for the model parameters $S=1/2$, $U=12t$, $\jh=0.2U$,
$J=4t^2/\Delta_0=0.2\bar{7}t$, and $\so=0.2t$ (the same model parameters as in Fig. 2 in the main text)
at the small temperature $T=0.02t$ with the number of bath sites $n_b=6$ in the
ED impurity solver.
We have plotted the local magnetizations $m=|\la s_i^z \ra|$ and $M=|\la S_i^z \ra|$
and the local Kondo correlation $K=\la \vec{S}_i \cdot \vec{s}_i\ra$ in Fig. \ref{fig:T0.02}(a),
the individual spin contributions to the Hall
conductance $\sigma_{yx}^\alpha$ in Fig. \ref{fig:T0.02}(b), and the charge gap for spin $\up$ and $\down$ in Fig. \ref{fig:T0.02}(c)
vs the alternating sublattice potential $\delta$. The charge gap is extracted from the local spectral function plotted
vs frequency in Fig. \ref{fig:T0.02}(d) for different values of $\delta$ across the topological phase transitions
$\delta_{c_1} \simeq 6.2t$ and $\delta_{c_2} \simeq 8.2t$.
The red vertical dashed lines in Fig. \ref{fig:T0.02}(b) denote the prediction of
the topological Hamiltonian method for the topological phase transitions at $T=0$.

The results in Fig. \ref{fig:T0.02}(a) confirm that for small values of $\delta$
the itinerant spin $\vec{s}_i$ and the localized spin
$ \vec{S}_i$ at each lattice site are in the triplet state corresponding to
$K=\la \vec{S}_i \cdot \vec{s}_i\ra \to 1/4$.
The low-energy properties of the system can effectively be described by the spin-1 Heisenberg model, i.e., $S=1/2$.
The magnetization of the system is due to the
Heisenberg interaction.
The N\'eel temperature $\tn \simeq 0.53t$ at $\delta=0$ in Fig. 2 in the main text nicely matches the
expected N\'eel temperature \cite{Hafez-Torbati2022} of the $\mathcal{S}_{\rm tot}=1$ Heisenberg model
$\tn=ZJ\mathcal{S}(\mathcal{S}+1)/3= 0.5\bar{5}t$  with the coordination number $Z=3$.
For large values of $\delta$ the magnetic properties of the system is mainly due to the localized
spins. The itinerant electrons show a very weak magnetization.
The N\'eel temperature for large values of $\delta$ in Fig. 2 in the main text
nicely approaches the expected N\'eel temperature of the $\mathcal{S}_{\rm tot}=1/2$ Heisenberg model
$\tn = 3J/4 \simeq 0.21t$ \cite{Hafez-Torbati2022}.

The Hall conductance $\sigma_{yx}^\alpha$ for the spin component $\alpha$ in Fig. \ref{fig:T0.02}(b) unveils
the emergence of the AFCI at intermediate values of $\delta$.
The results correspond to the solution with the magnetization on the higher-energy sublattice
pointing in the positive $z$ direction.
The location of the transition points nicely
matches with the ones predicted by the topological Hamiltonian approach for $T=0$ (the red vertical
dashed lines).
Figure \ref{fig:T0.02}(c) confirms that the topological phase transitions for $\alpha=\up$ in
Fig. \ref{fig:T0.02}(b) is accompanied by the charge gap closing.

\begin{figure}[t]
   \begin{center}
   \includegraphics[width=0.36\textwidth,angle=-90]{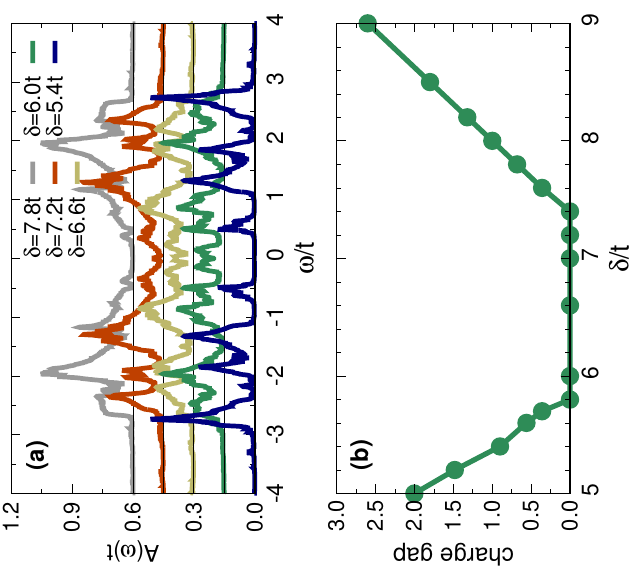}
   \caption{(a) Local spectral function vs frequency near the Fermi energy $\omega=0$ for different values
   of the alternating sublattice potential $\delta$. (b) The charge gap vs $\delta$.
   The results are for the model parameters $S=1/2$, $U=12t$, $\jh=0.2U$, $J=4t^2/\Delta_0=0.2\bar{7}t$,
   and $\so=0.2t$ at the paramagnetic temperature $T=0.6t$. The number of bath sites $n_b=6$ is used
   in the ED impurity solver.}
   \label{fig:T0.6}
   \end{center}
\end{figure}

The metallic phase separating the Mott and the band insulators at high temperatures in Fig. 2 in the main text
can be understood from the charge excitations depicted in Fig. \ref{fig:T0.6}. The results are for
$S=1/2$, $U=12t$, $\jh=0.2U$, $J=4t^2/\Delta_0=0.2\bar{7}t$, and $\so=0.2t$ (the same model parameters as in Fig. 2
in the main text) at the temperature $T=0.6t$. The number of bath sites $n_b=6$ is used in the ED impurity solver.

Fig. \ref{fig:T0.6}(a) displays the local spectral function vs frequency near the Fermi energy $\omega=0$ for
different values of $\delta$. One can see the finite spectral weight developing at the Fermi energy for the intermediate
values of $\delta$. The charge gap extracted from the local spectral function is plotted vs $\delta$ in
Fig. \ref{fig:T0.6}(b). In the Mott insulator regime for small values of $\delta$ the charge gap decreases upon
increasing $\delta$. It closes for a finite interval of $\delta$ and then increases upon increasing
$\delta$ characteristic of the band insulator phase.

\section{Mean-field N\'eel temperature of the effective spin model}
In the limit $(\Delta_0 -2\delta) \gg t,\so$ the generalized Kondo lattice model, in Eq. (2) in the main text,
can be mapped by a unitary transformation \cite{Fazekas1999} to the effective low-energy spin model
\begin{align}
\label{eq:spin}
 H&=J_{\rm eff} \sum_{\langle i,j\rangle} \vec{s}^{\vpdag}_{i} \cdot \vec{s}^{\vpdag}_{j}
 +J'_{\rm eff} \sum_{[ i,j]} \left( {s}^{z}_{i} {s}^{z}_{j}-{s}^{x}_{i}  {s}^{x}_{j}-{s}^{y}_{i}  {s}^{y}_{j} \right)
  \nn \\
-&2J_{\rm H} \sum_{i} \vec{s}_i \cdot \vec{S}_i
+J\sum_{\langle i,j\rangle} (\vec{{S}}_i \cdot \vec{{S}}_j
+\vec{{S}}_i \cdot \vec{{s}}_j
+\vec{{s}}_i \cdot \vec{{S}}_j) \ ,
\end{align}
where $J_{\rm eff}=4t^2\Delta_0/(\Delta_0^2-4\delta^2)$ and $J'_{\rm eff}=4\so^2/\Delta_0$. A mean-field decoupling
of the non-local Heisenberg interactions simplifies the Hamiltonian in Eq. \eqref{eq:spin} to
\begin{equation}
\label{eq:spin:mf}
 H_{\rm MF}=-\sum_{i} h_{i}^{\rm iti} s^z_{i} -\sum_{i} h_{i}^{\rm loc} S^z_{i} -2J_{\rm H} \sum_{i} \vec{s}_i \cdot \vec{S}_i
\end{equation}
where we have assumed a N\'eel AF order in the $z$ direction. The effective magnetic fields acting on the itinerant and
the localized spins are given by
\begin{subequations}
\begin{align}
 h_{i}^{\rm iti} &:=   (Z J_{\rm eff} - Z' J'_{\rm eff}) \langle s^z_{i} \rangle +Z J \langle S^z_{i} \rangle  \ ,  \\
h_{i}^{\rm loc} &:= Z J \langle s^z_{i} +S^z_{i} \rangle  \ ,
\end{align}
 \end{subequations}
 where $Z=3$ and $Z'=6$ denote the first and the second neighbor coordination numbers, respectively.
 We consider the localized spin $S=1/2$.
 Straightforward calculations for the local magnetizations $\langle s^z_{i}\rangle$ and  $\langle S^z_{i} \rangle$
 of the mean-field Hamiltonian \eqref{eq:spin:mf} at  the inverse temperature $\beta=1/(k_{\rm B}T)$ lead to the expressions
 \begin{subequations}
 \label{eq:mf}
  \begin{align}
   \langle s^z_{i}\rangle =& \frac{1}{2} \frac{\sinh(\beta h_i^+) + \kappa e^{-\beta \jh}\sinh(\beta h_i^-/\kappa)}{\cosh(\beta h_i^+)+\cosh(\beta h_i^-/\kappa)}  \ , \\
   \langle S^z_{i}\rangle=& \frac{1}{2} \frac{\sinh(\beta h_i^+) - \kappa e^{-\beta \jh}\sinh(\beta h_i^-/\kappa)}{\cosh(\beta h_i^+)+\cosh(\beta h_i^-/\kappa)} \ ,
  \end{align}
 \end{subequations}
where $h_i^-:=(h_{i}^{\rm iti} - h_{i}^{\rm loc})/2$, $h_i^+:=(h_{i}^{\rm iti} + h_{i}^{\rm loc})/2$, and
$\kappa:=h_i^-/\sqrt{h_i^-{}^2+\jh^2}$.

The iterative solution of the mean-field equations \eqref{eq:mf} determines the temperature dependence
 of the local magnetizations $\langle s^z_{i}\rangle$ and  $\langle S^z_{i} \rangle$. The obtained N\'eel
 temperature  $\tilde{T}_{\rm N}$ is compared with the N\'eel temperature  $\tn$ of the generalized Kondo
 lattice model in Fig. 2 in the main text.

\begin{widetext}

\section{Chiral edge states}
To study the chiral edge states we consider a cylindrical geometry with open boundary conditions in $x$
and periodic boundary conditions in $y$ directions, as sketched in Fig. 4 in the main text. After a
Fourier transform in the $y$ direction we obtain the Bloch Hamiltonian
\begin{subequations}
\label{eq:Hk}
\begin{gather}
 H^{(0)}(k_y)=H_{t}(k_y)+H_{\rm so}(k_y)+H_{\delta}(k_y)+H_{\mu}(k_y) \ , \\
 H_{t}(k_y)=+t\sum_{\alpha=\up,\down} \sum_{x=0}^{N_x-1} \left( e^{{\rm i}k_ya(-1)^x} c^{\dagg}_{x,0;\alpha} c^{\vpdag}_{x,1;\alpha}
 +\sum_{y=0}^1 c^{\dagg}_{x+1,y;\alpha} c^{\vpdag}_{x,y;\alpha}
 \right) +{\rm H.c.} \ , \\
 H_{\rm so}(k_y)=-{\rm i}\so \sum_{\alpha=\up,\down} \sum_{x=0}^{N_x-1}\sum_{y=0}^1 \sigma_{\!\alpha \alpha}^z (-1)^{x+y}
 \left(
 c^{\dagg}_{x+2,y;\alpha} c^{\vpdag}_{x,y;\alpha} -2\cos(k_y a)c^{\dagg}_{x,y+1;\alpha} c^{\vpdag}_{x+1,y;\alpha}
 \right)+{\rm H.c.} \ , \\
 H_{\delta}(k_y)+H_{\mu}(k_y)= \sum_{\alpha=\up,\down} \sum_{x=0}^{N_x-1}\sum_{y=0}^1
 \left(
 (-1)^{x+y+1}\delta -\frac{1}{2}(-1)^{x+y+1}\la S_{\rm odd}^z \ra Z_{x}{\rm sgn}(\alpha) J -\frac{U}{2} \right) c^{\dagg}_{x,y;\alpha} c^{\vpdag}_{x,y;\alpha}
\end{gather}
\end{subequations}
\end{widetext}
where $(x,y)$ specifies a lattice site in the unit cell with $x=0,1,\cdots,N_x-1$ and $y=0,1$.
The dependence of $c^{\dagg}_{x,y;\alpha}$ on the momentum $k_y$ is implicit, and
$c^{\dagg}_{x,2;\alpha}\equiv c^{\dagg}_{x,0;\alpha}$. The parameter $a$
represents the distance between the NN sites.
Note that the distance $2a$ is used as the unit of length in the plot in Fig. 5 in the main text.
The NN coordination number $Z_{x}$ equals 2 for $x$ at the edges and equals $3$ for $x$ in the bulk.
Assuming the uniform N\'eel AF order, as illustrated in Section \ref{sec:dmft}, one has
$\la S_{x,y}^z \ra=(-1)^{x+y+1}\la S_{\rm odd}^z\ra$. It is supposed that $\la S_{\rm odd}^z\ra$ is
already determined from the periodic boundary conditions in both directions. The Bloch Hamiltonian \eqref{eq:Hk} can be seen
as an effective two-leg ladder model involving hoppings up to the third neighbor.

The  Green's function ${\bf G}_\alpha(\omega+{\rm i}\eta,k_y)$ is found using
the matrix representation of the Bloch Hamiltonian \eqref{eq:Hk} and the DMFT self-energy $\bs{\Sigma}_\alpha(\omega+{\rm i}\eta)$.
The momentum-resolved spectral function $A_{\alpha,\vec{d}}~(\omega,k_y)$ for the spin component $\alpha$ at the lattice site
$\vec{d}=(x,y)$ in the unit cell is then easily obtained via Eq. 4 in the main text.
We have used the broadening factor $\eta=0.01t$ as mentioned in the main text.


%